\documentclass[printer]{aa}

\usepackage{natbib}
\usepackage{amsmath}
\usepackage{graphicx}
\usepackage{txfonts}

\newcommand{\rg}{GM_{\mathrm{BH}}/c^2}
\newcommand{\rgc}{GM_{\mathrm{BH}}/c^3}
\newcommand{\rgd}{GM_{\rm BH}/c^2D}
\newcommand{\tu}{GM_{\mathrm{BH}}/c^3}
\newcommand{\trat}{T_{\rm p}/T_{\rm e}}

\newcommand{\msun}{M_{\rm \odot}}
\newcommand{\mbh}{M_{\rm BH}}
\newcommand{\mdot}{\dot{M}}
\newcommand{\mdotu}{\rm M_\odot yr^{-1}}

\usepackage{color}

\begin{document}

%\title{Modeling the supermassive black hole in M87 with GRMHD simulations of a  radiatively inefficient accretion flow with a relativistic jet}
\title{GRMHD simulations of the jet in M87}
   \author{Mo{\'s}cibrodzka M., Falcke H., Shiokawa H.}
   \institute{Department of Astrophysics/IMAPP, Radboud University,
     P.O. Box 9010, 6500 GL Nijmegen, The Netherlands\\
     \email{m.moscibrodzka@astro.ru.nl}}
   \date{Received May, 2015; accepted x, 2015}
\titlerunning{GRMHD jet in M87}
\authorrunning{Mo{\'s}cibrodzka, Falcke, Shiokawa}
\author{Monika Mo{\'s}cibrodzka~\inst{1}, Heino Falcke~\inst{1,2}, Hotaka
  Shiokawa~\inst{3\thanks{now moved to Harvard-Smithsonian Center for
      Astrophysics, 60 Garden Street, Cambridge, MA 02138, USA}}}
\institute{$^1$Department of Astrophysics/IMAPP,Radboud University
  Nijmegen,P.O. Box 9010, 6500 GL Nijmegen, The Netherlands\\
$^2$ASTRON, Oude Hoogeveensedijk 4, 7991 PD, Dwingeloo, The Netherlands\\
$^3$Department of Physics \& Astronomy, The Johns Hopkins University 3400 N. Charles Street
Baltimore, MD 21218, USA\\
\email{m.moscibrodzka@astro.ru.nl}}
\date{Received May, 2015; accepted x, 2015}

\abstract
    { The connection between black hole, accretion disk, and radio jet can be constrained
     best by fitting models to observations of nearby low-luminosity galactic nuclei, in particular the well-studied sources Sgr~A* and
     M87. There has been considerable progress in modeling the central
       engine of active galactic nuclei by an accreting supermassive black
       hole coupled to a relativistic plasma jet.  However, can a single model
     be applied to a range of black hole masses and accretion rates?  }
   { Here we want to compare the latest three-dimensional numerical model, originally 
       developed for Sgr A* in the center of the Milky Way, to radio
       observations of the much more powerful and more massive black hole in
       M87. }
   { 
     We postprocess three-dimensional GRMHD models of a jet-producing radiatively inefficient
     accretion flow around a spinning black hole using 
     relativistic radiative transfer and ray-tracing
     to produce model spectra and images. As a key new ingredient in these models, we allow the proton-electron
     coupling in these simulations depend on the magnetic properties of the plasma.
        }
   {
   We find that the radio emission in M87 is described well by a
   combination of a two-temperature accretion flow and a hot
   single-temperature jet. Most of the radio emission in our
   simulations comes from the jet sheath. The model fits the basic
   observed characteristics of the M87 radio core: it is
   "edge-brightened", starts subluminally, has a flat spectrum, and
   increases in size with wavelength.  The best fit model has a mass-accretion rate of
   $\mdot \sim 9\times 10^{-3} \, \mdotu$ and a total jet power of $P_{\rm j} \sim 10^{43}
   \, \mathrm{erg\,s^{-1}}$. Emission at 
     $\lambda=1.3\,\mathrm{mm}$ is produced by the counter-jet
     close to the event horizon. Its characteristic crescent shape surrounding the black hole shadow
     could be resolved by future millimeter-wave VLBI experiments.}
      {
       The model was successfully derived from one for the supermassive
       black hole in the center of the Milky Way by appropriately scaling
       mass and accretion rate. This suggests the possibility that this model
       could also apply to a wider range of low-luminosity black holes.
       }
    \keywords{ Accretion, accretion disks --  Black hole physics --  Relativistic processes -- 
Galaxies: jets --  Galaxies: nuclei }
 \maketitle

\section{Introduction}

The most notable signature of an active black hole (BH) is a radio jet, but
the exact processes of jet production 
by an accretion disk around a spinning BH, as well as its
collimation and acceleration, has not been fully established.  The 
radio core in the center of the Milky Way (hereafter Sgr~A*) and 
in the center of the massive elliptical galaxy M87 --- two sources that can be
observed with unprecedented resolution --- display nearly flat radio spectra
that are characteristic of relativistic jets 
(\citealt{blandford:1979}; \citealt{falcke:1995}). For
both objects, the combination of the putative central BH mass ($\mbh$)
and distance ($D$) provide an expected angular size of the BH on
the sky ($2\sqrt{27} (\rgd) \approx 54$ and $38 \mathrm{\mu as}$, for Sgr~A*
and M87, respectively). In principle, therefore, they can be resolved (together
with the surrounding plasma) with current radio and millimeter very long
baseline interferometers (VLBI). Millimeter-VLBI observations promise to
provide essential clues about the jet-disk-BH connection because they will probe
plasma in the immediate vicinity of a BH in these two sources. Since Sgr~A*'s
intrinsic geometry is smeared out by an interstellar scattering screen
(e.g., \citealt{bower:2014}), the nature of the radio emission near the central
supermassive black hole (SMBH) remains hidden, so it is somewhat difficult to
probe its putative jet intrinsic geometry and properties. On the other hand,
the radio core of M87 is resolved into a clear jet structure and can be readily used as a
laboratory for testing various theoretical models of the magnetized plasma
flows onto compact objects.

The M87 core is a prototype of a radio loud low-luminosity active galactic
nuclei (\citealt{ho:2008}, and references therein). The bolometric luminosity
of the core is estimated to be $L/L_{\rm Edd} \approx 10^{-7}$, which suggests
that the powerful (the jet power is $P_j\,\approx\,10^{45}
\mathrm{erg\,s^{-1}}$, \citealt{gasperin:2012}) kpc-scale jet emerges from a
radiatively inefficient accretion flow (RIAF) onto the supermassive BH
\citep{yuan:2014}. A similar, downsized model is often used to explain the $L/L_{\rm
  Edd} \approx 10^{-9}$ emission from Sgr~A*.
 In fact, radio cores seem to be rather ubiquitous in low-luminosity AGN
  (LLAGN) \citep{nagar:2000} and scale with mass and accretion rate in a
  simple way \citep{merloni:2003,kording:2006}. One would therefore expect a generic RIAF+jet model to equally scale
  between AGN of different masses and accretion rates.

Advances in the field of numerical astrophysics now allows
magnetized RIAFs and their jets to be simulated almost from the first principles (e.g.,
\citealt{abramowicz:2013}; \citealt{yuan:2014}; and references therein). 
General relativistic radiative transfer (GRRT) models can predict the 
 general relativistic magnetohydrodynamical (GRMHD)
simulation spectrum and appearance. This allows us to compare dynamical
models directly to VLBI observations (e.g., \citealt{dexter:2012};
\citealt{hilburn:2012}; \citealt{moscibrodzka:2014}).

However, a few crucial uncertainties in the GRMHD simulations and GR radiative
transfer simulations do not yet allow one to tightly constrain these models and to
closely examine whether the BH spin or the magnetic fields play the most critical role
in jet formation. One of the main uncertainties in GRMHD simulation is
associated with the distribution function (DF) of radiating particles
(electrons). Electron DF is not explicitly computed in any of the current
GRMHD simulations. Moreover, as shown in \citet{moscibrodzka:2014}, the
appearance of a GRMHD model strongly depends on the details of the assumed
electron temperature (the so-called "painting" of GRMHD
simulations). Moreover, the true electron DF in the magnetized plasma can vary
with space and time, which leaves us many degrees of freedom in the
interpretations of observational data.

Recently, we have phenomenologically found a simplified but {\it natural},
location-dependent prescription for electron DFs in the GRMHD models that is
able to reproduce observational characteristics of Sgr~A*
(\citealt{moscibrodzka:2013}; \citealt{moscibrodzka:2014}). In particular, the
flat spectrum of the source can be reproduced when we assume that electrons
along the jet funnel (i.e., in the jet sheath) produced by the RIAF are hotter than
those in the RIAF (accretion disk) itself. In this electron model, we assumed
that electrons in the accretion disk have a thermal, Maxwellian distribution
function, but they are weakly coupled to protons, so the plasma is
two-temperature in the disk, and electrons are rather cool. At the same time,
to account for radio emission of Sgr~A*, we assumed that the electrons in the
jet and in the jet wall are much hotter than in the accretion
disk. Under this assumption, the strongly magnetized jet becomes brighter than
the relatively weakly magnetized disk in radio wavelengths. In other words,
the model appears to the observers as a jet-dominated advective system (JDAF). 

We have also found that the emission from the tenuous jet wall in the
simulations can account for the entire radio spectrum of Sgr~A* (when the jet
extends by 2-3 orders of magnitude in radius) and the
two-temperature disk is visible around millimeter wavelengths and shorter. The jet images were
edge-brightened in a similar way because it is observed in the M87 jet (e.g., \citealt{ly_walker:2007}). 
The
next logical step is thus to scale the same model to M87 to test if the model can
reproduce the radio observations of the M87 jet. Furthermore, the M87 core is also a
target of the Event Horizon Telescope (EHT, \citealt{doeleman:2012}), which is
a millimeter Very Long Baseline Interferometric experiment to image the central
BH and its surrounding in Sgr~A* and M87. It is natural to ask
whether the BH illuminated by the footprint of a relativistic jet is
detectable by the EHT \citep{dexter:2012}.

The goal of this work is to examine how the footprint of the jet would appear
in radio observations and to study how the apparent size of the jet depends
on wavelength.  Ultimately, the model should be constrained by the
multi-wavelength VLBI data at 7\,mm, 3.5\,mm, and 1.3\,mm, which is already available.
In this work, we use the same three-dimensional (3-D) GRMHD model that was used by
us to model Sgr~A* \citep{moscibrodzka:2014}, but rescaled to 
the mass of the BH in M87.

The paper is organized as follows. In Sect.~\ref{sec:model}, we briefly
present the 3-D GRMHD model and radiative transfer technique and describe how
we scale the model to the M87 system. We also introduce some of the observational
constraints used in our models.  In Sect.~\ref{sec:results}, we compare the
model spectra and radio images directly to $\lambda=$7, 3.5, and 1.3\,mm
observations of the inner jet in M87. We discuss our results in the context of
results found in the literature in Sect.~\ref{sec:discussion}.  We conclude in
Sect.~\ref{sec:conclusions}.

\section{Models of a jet}\label{sec:model}

\subsection{Dynamics of plasma and magnetic fields}

Our time-dependent model of a radiatively inefficient accretion flow onto a
BH is based on the fully, three-dimensional (3-D) GRMHD simulation
carried out by \citet{shiokawa:2013} (run b0-high in Table 5.1 in
\citealt{shiokawa:2013}).  The simulation started from a torus in hydrodynamic
equilibrium in the equatorial orbit around a rotating BH
\citep{fishbone:1976}. The torus initially had a pressure maximum at $24\,\rg$
and an inner edge at $12\,\rg$. It was seeded with a weak poloidal field that
follows the isodensity contours (single-loop model, see
\citealt{gammie:2003}). The dimensionless BH spin was
$a_*\simeq\,0.94$. The corresponding radius of the event horizon was
$r_{\mathrm{h}}=\,1.348\,\rg$ and the innermost stable circular orbit (ISCO)
was located at $r_{\mathrm{ISCO}}=\,2.044\,\rg$.  The inner boundary of the
computational domain was just inside the event horizon and the outer boundary
was at $R_{\mathrm{out}} = 240\,\rg$. The model was evolved for $14~000\,\rgc$,
which is equivalent to about 19 orbital periods at $r=24\,\rg$.

Jets are naturally produced in the GRMHD simulation and are defined as the strongly magnetized
and nearly empty regions above the BH poles. We refer to this region
as the jet spine. Close to the BH, most of the energy of the jet
spine is stored in the magnetic fields that cannot be radiated
away efficiently.  Although it is uncertain, the plasma density in the jet
spine is most likely very low; as a result, this region would not produce any significant
radio or synchrotron emission (see, e.g., \citealt{moscibrodzka:2011}).  However,
as noted by \citet{moscibrodzka:2013} and \citet{moscibrodzka:2014}, the jet
definition should also include the jet sheath, which is a tenuous layer of gas that moves
along the jet spine. The jet sheath is less magnetized in comparison to the
spine (the plasma $\beta$ parameter decreases with radius from 50 to 1 in the
jet sheath, while $\beta \lesssim 1$ in the spine), but has higher matter
content that can be constantly resupplied by an accretion disk.

\subsection{Emission model}

The radiative properties of the dynamical model are studied by post-processing
the GRMHD model with radiative transfer (RT) computations.  Calculations are
carried out using the same tools as in \citet{moscibrodzka:2009}. The spectral
energy distributions (SEDs) are computed using a Monte Carlo code for
relativistic radiative transfer {\tt grmonty} \citep{dolence:2009} that
includes radiative processes, such as synchrotron emission, self-absorption,
and inverse-Compton processes.  To create model images, a ray-tracing
radiative transfer scheme is used \citep{noble:2007}. Both RT schemes solve RT
equations for total (unpolarized) intensity along null-geodesics trajectories.

Local plasma synchrotron emissivity ($j_\nu$) and absorptivity ($\alpha_\nu$)
depend on the assumed electron distribution function, the angle between magnetic field
and the plasma velocity vectors, the magnetic field strength, and the electron number
density. The resulting spectrum detected by an observer will therefore depend on
the observer's orientation with respect to the synchrotron source, structure, and
bulk speed of plasma and magnetic field geometries.  Near a BH, the
emission is also affected by spacetime curvature.  In our models, we take all these effects
into account; i.e., the radiative transfer code includes all the
variables mentioned earlier, except the electron distribution function (see next
section), directly from the GRMHD accretion flow model.

\subsection{Electron DF in the disk and in jets}

It is almost certain that electron DFs are non-thermal, power-law functions,
but for simplicity, in our modeling, electrons are described by
a Maxwell-J{\"u}ttner distribution parameterized by
$\Theta_{\mathrm{e}}=kT_{\mathrm{e}}/m_{\mathrm{e}}c^2$.  While the proton
temperature $T_{\mathrm{p}}$ is provided by the GRMHD simulation, we assume
that the electron temperature $T_{\mathrm{e}}$ depends on plasma
magnetization.  The electron temperature is calculated from the following
formula:
\begin{equation}\label{eq:te}
\frac{T_{\mathrm{p}}}{T_{\mathrm{e}}}=R_{\mathrm{high}} \frac{b^2}{1+b^2}  +
R_{\mathrm{low}} \frac{1}{1+b^2}
\end{equation}
where $b=\beta/\beta_{\mathrm{crit}}$,
$\beta=P_{\rm{gas}}/P_{\mathrm{mag}}$, and $P_{\mathrm{mag}}= B^2/2$.  We
assume that $\beta_{\mathrm{crit}}=1$, and $R_{\mathrm{high}}$ and
$R_{\mathrm{low}}$ are temperature ratios that describe the electron-to-proton
coupling in the weakly magnetized (disk, high $\beta$ regions) and strongly
magnetized regions (jet, low $\beta$ regions), respectively.
In Eq.~\ref{eq:te}, the proton-to-electron temperature ratio scales with
$\beta^2$, which guarantees that the regions of strong and weak proton-to-electron
coupling are clearly defined, and their radiations are easy to distinguish and
to interpret, which would not be the case if the temperature ratio scaled linearly
with $\beta$. In our RT model,
the accurate synchrotron emissivities for thermal, relativistic electrons are
adopted from \citet{leung:2011}.

Our new electron temperature definition (Eq.~\ref{eq:te}), which describes the
electron temperatures in the jet and disk zones, has been slightly modified
compared to the one used in \citealt{moscibrodzka:2014} (in which we defined the
jet zones as unbound plasma, and electrons in the jet had constant
temperature).  This is done to avoid artifacts such as sharp boundaries
between the disk and jet zones. Nevertheless, the current model is similar to
the one in \citet{moscibrodzka:2014}, where the accretion disk and the jets
are described as a two-temperature and a single-temperature plasma,
respectively. Here, we simply associate the plasma temperatures with plasma
magnetization, which is physically more intuitive.

\begin{figure*}
 \begin{center}
 \includegraphics[trim=3.9cm 0.5cm 3.9cm 0.5cm,clip=true,width=0.25\textwidth]{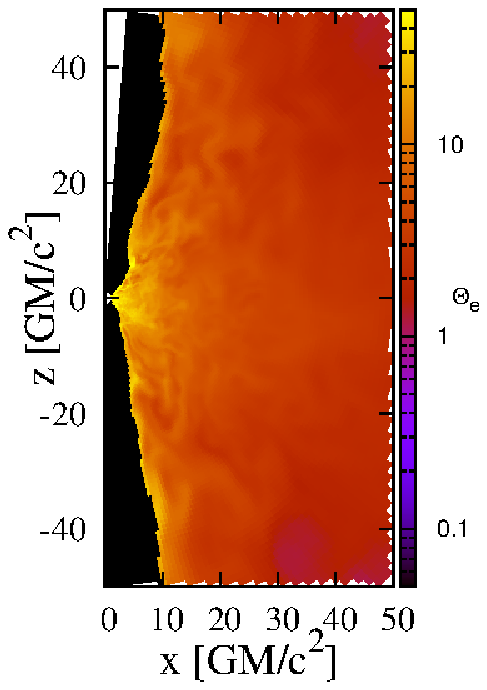}
 \includegraphics[trim=3.9cm 0.5cm 3.9cm 0.5cm,clip=true,width=0.25\textwidth]{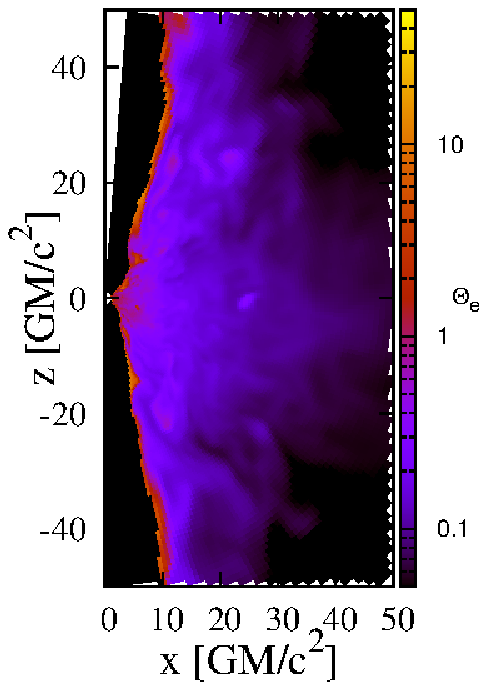}
 \includegraphics[trim=3.9cm 0.5cm 3.9cm 0.5cm,clip=true,width=0.25\textwidth]{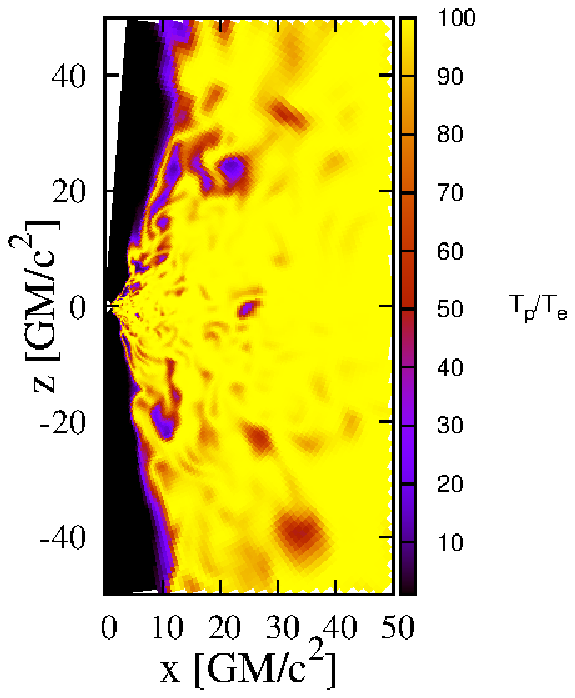}
 \caption{Dimensionless electron temperature, $\Theta_{\mathrm{e}}=kT_{\mathrm{e}}/m_{\mathrm{e}}c^2$, in
   the model with $R_{\mathrm{high}}=1$ (left panel) and $R_{\mathrm{high}}=100$
   (middle panel). The right panel shows the proton-to-electron temperature
   ratio in the model with $R_{\mathrm{high}}=100$. 
 The maps show the slices through the 3-D GRMHD model (run b0-high in Table 5.1
 in \citealt{shiokawa:2013}) along the BH spin axis.}\label{fig:thetae_xz}
 \end{center}
 \end{figure*}

\begin{figure*}
 \begin{center}
 \includegraphics[trim=1.5cm 0cm 1.5cm 0cm,clip=true,width=0.25\textwidth]{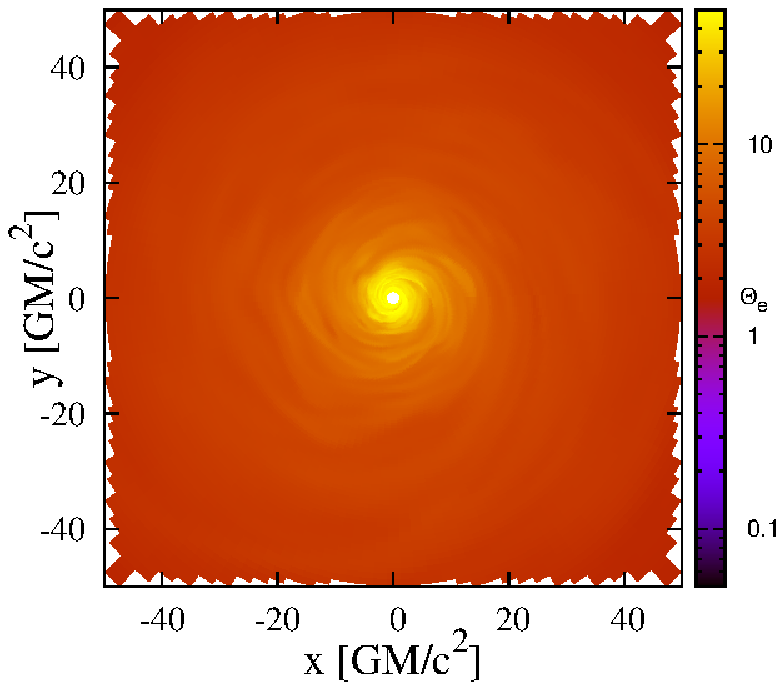}
 \includegraphics[trim=1.5cm 0cm 1.5cm 0cm,clip=true,width=0.25\textwidth]{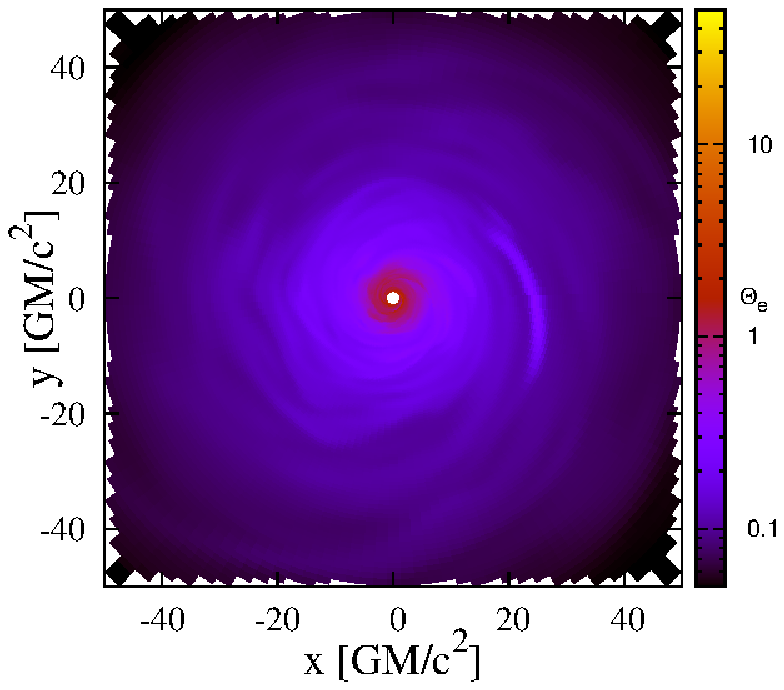}
 \includegraphics[trim=1.5cm 0cm 1.5cm 0cm,clip=true,width=0.25\textwidth]{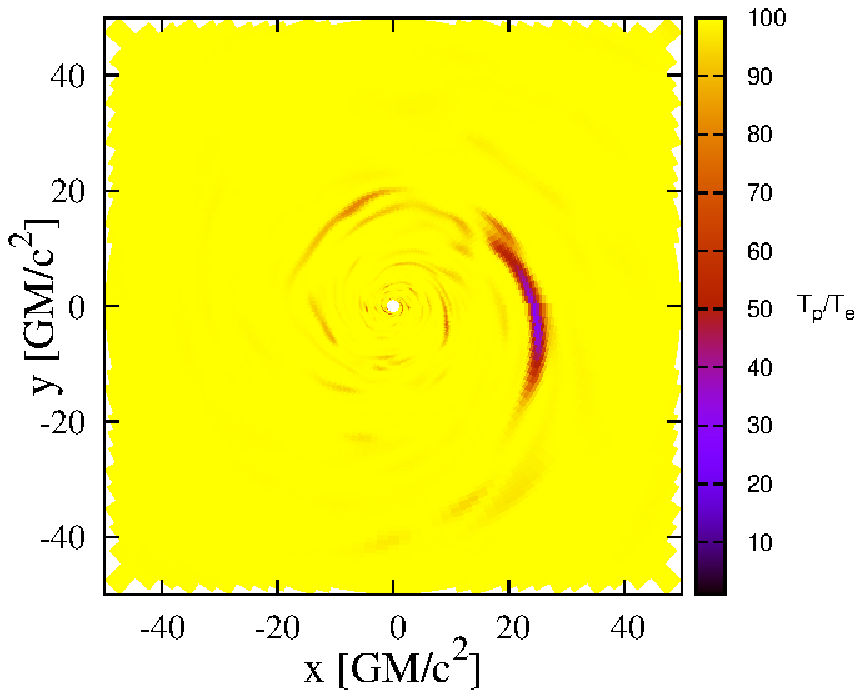}
 \caption{Same as in Fig.~\ref{fig:thetae_xz}, 
 but for the slices through the disk equatorial plane.}\label{fig:thetae_xy}
 \end{center}
 \end{figure*}

We consider six models with fixed $R_{\rm low}=1$ and varying $R_{\rm
  high}=1,5,10,20,40,100$.  In Figs.~\ref{fig:thetae_xz} and~\ref{fig:thetae_xy}, we show maps of electron temperature calculated using
Eq.~\ref{eq:te} for two extreme cases ($R_{\rm high}$=1 and 100) and the
proton-to-electron temperature ratio when $R_{\rm high}$=100. The
electron temperature is expressed in units of electron rest mass, $\Theta_{\rm
  e}=kT_{\rm e}/m_{\rm e}c^2$, and $\Theta_{\rm e} \gtrsim 0.1$ for plasma to
emit synchrotron radiation. In models with $R_{\rm high}$=100, the plasma with
temperatures $\Theta_{\rm e} > 1$ occupies the tenuous jet wall (see
Fig.~\ref{fig:thetae_xz}, right panel). 

For fixed $R_{\rm low}=1$ and  $R_{\rm high}>100,$ the electron temperature in the
jet wall becomes sub-relativistic and does not produce any continuum synchrotron emission. 
Values of $R_{\rm low}$ that are less than unity are also physically possible when 
electrons are additionally heated up, such as in magnetic field reconnection
events or by turbulence. Our models assume ideal-MHD conditions, and turbulence in the tenous 
jet wall is unresolved. For the self-consistency of models, 
we do not consider models with $R_{\rm low}<1$.


\subsection{Scaling the model to M87 core}

The core and jet of M87 have been observed mostly in radio but also in optical
and X-ray bands. An excellent overview of M87 jet properties and a list of
references to the source observations is given in \citet{biretta_proc:1995}
and more recently in \citet{nakamura:2013}.

To scale the dimensionless GRMHD simulation to astrophysical sources, one has to
provide the central BH mass, its distance to the observer, and the mass
of the accretion disk around the BH (which is equivalent to changing
the mass accretion rate onto the SMBH).  Based on the stellar or gas dynamics
in the core, the mass of the M87 central BH has been estimated as
$M_{\rm BH}=3.2, 3.5$, and $6.2 \times 10^9 {\mathrm{M_\odot}}$ (by
\citealt{macchetto:1997}; \citealt{gebhardt:2011}; \citealt{walsh:2013},
respectively).  Here, we adopt the most recent value from the observations,
$M_{\mathrm{BH}}=6.2 \times 10^{9} {\mathrm{M_\odot}}$, in all of our
models.  The distance to M87 is assumed to be the same as the mean distance
to the Virgo cluster: $D$=16.7 Mpc \citep{mei:2007}.

The BH mass $M_{\mathrm{BH}}$ sets the length unit (the gravitational
radius) $\rg=9.2 \times 10^{14}\,{\mathrm{cm}} = 2.89\times 10^{-4}$ pc and
sets the model time unit to be $\tu=8.5\,h$.  The GRMHD snapshots used in the
radiative transfer modeling were evolved for 10~000 $\rgc$ time units, which
corresponds to approximately 10 yr.  We analyzed the GRMHD simulation time
slices for a short time interval of about $1000\,\rgc\equiv 350\,d$ dumped by
the code every $10\,\rgc\equiv 3.5\,d$.  The GRMHD model diameter of
$480\,\rg$ corresponds to about 1.8 mas on the sky for the chosen
$M_{\mathrm{BH}}$ and $D$.

The orientation of the source with respect to the observer is described by two angles:
inclination angle, $i$, and position angle, $\mathrm{PA}$.  These can be
estimated from the observations of the M87 jet at radio and optical
wavelengths.  However, we simply assume two inclination (viewing) angles,
$i=20$ and $160\degr$. At the viewing angle of $160\degr$, the model rotates in
the opposite direction to $i=20\degr$. The direction of the rotation of
the system is also examined. In the models presented in this paper, we
adopted the position angle of the BH spin on the sky as $PA=290\degr$,
estimated from radio observations by \citet{reid_1.6:1982}.

The mass accretion rate onto the SMBH, $\mdot$, is a free parameter.  The
accretion rate can be estimated by fitting the model SED to the
multiwavelength observational data points.

The X-ray luminosity of the central region (<2'') around the BH is
estimated as $4-7 \times 10^{40}\,{\mathrm{erg\,s^{-1}}}$
(\citealt{marshall_xray:2002}; \citealt{dimatteo:2003}).  The radio spectrum
is nearly flat with the flux $F_{\mathrm{R}} \sim 1\, \mathrm{Jy}$. Our SED
models, computed for a range of $R_{\mathrm{high}}$ values, are normalized
such that the flux at $\lambda$=1.3\,mm is 1 Jy (see
\citealt{doeleman:2012}).  The normalization can be achieved by adjusting
$\mdot$. In the models presented in this work, $\mdot$ ranges from $10^{-4}$
to $10^{-2} \,\mdotu$, and $R_{\mathrm{high}}$ from 1 to 100.

The mass accretion rate in the model is changed by multiplying the plasma
density by a constant number. The magnetic field strength in the
model is described by a dimensionless plasma parameter $\beta$, therefore by
changing the model density normalization, we automatically change the strength
of the magnetic field in the entire model ($B \sim n_e^{1/2}$).  The
observational constraint on the accretion rate can also be inferred from the
observed Faraday rotation at $\lambda$=1.3\,mm, and
${\mathrm{RM}}=\pm{\mathrm{a\,few}} \times 10^5\,\mathrm{rad\, m^{-2}}$
\citep{kuo:2014}. However, the $\dot{M}$ derived based on RM is strongly model
dependent. We calculate RM self-consistently based directly on the GRMHD
and RT models (see Sect.~\ref{dis:mdot}).

\section{Results}\label{sec:results}

\subsection{Modeling radiative efficiency and SEDs data}

\begin{table*}
\centering
\tiny
\caption{Summary of the radiative transfer models.}\label{tab:model_params}
\begin{tabular}{c c c c c c c c c}
\hline\hline
ID & $R_{\rm high}$ & $R_{\rm low}$ & $\beta_{\rm crit}$&$\mdot$ [$\mdotu$] &
$\epsilon_{\rm r}$~\tablefootmark{a}&$F_{\mathrm 7\,mm}$ [Jy]\tablefootmark{b} &
$F_{\mathrm 3.5\,mm}$ [Jy]\tablefootmark{c} & $F_{\mathrm 1.3\,mm}$
[Jy]\tablefootmark{d} \\
\hline
RH1 & 1   & 1 & 1&$1\times10^{-4}$ & 23   & 0.96 & 1.54 & 1 \\
RH5 & 5   & 1 & 1&$4\times10^{-4}$ & 5.1  & 0.37 & 0.71 & 1 \\
RH10 & 10  & 1 & 1&$1\times10^{-3}$ & 3.22 & 0.36 & 0.54 & 1\\
RH20 & 20  & 1 & 1&$3\times10^{-3}$ & 1.82 & 0.82 & 0.77 & 1\\
RH40 & 40  & 1 & 1&$5\times10^{-3}$ & 0.31 & 1.34 & 1.16 & 1\\
RH100 & 100 & 1 & 1&$9\times10^{-3}$ & 0.02 & 1.67 & 1.5  & 1\\
\hline
\end{tabular}
\tablefoot{
\tablefoottext{a}{But see \S~\ref{dis:cool} for discussion of radiative efficiencies.}\\
\tablefoottext{b}{FOV (camera field-of-view) = 480x480 $\rg$}\\
\tablefoottext{c}{FOV = 100x100 $\rg$}\\
\tablefoottext{d}{FOV = 40x40 $\rg$}
}
\end{table*}

\begin{figure*}
\begin{center}
\includegraphics[width=0.48\textwidth]{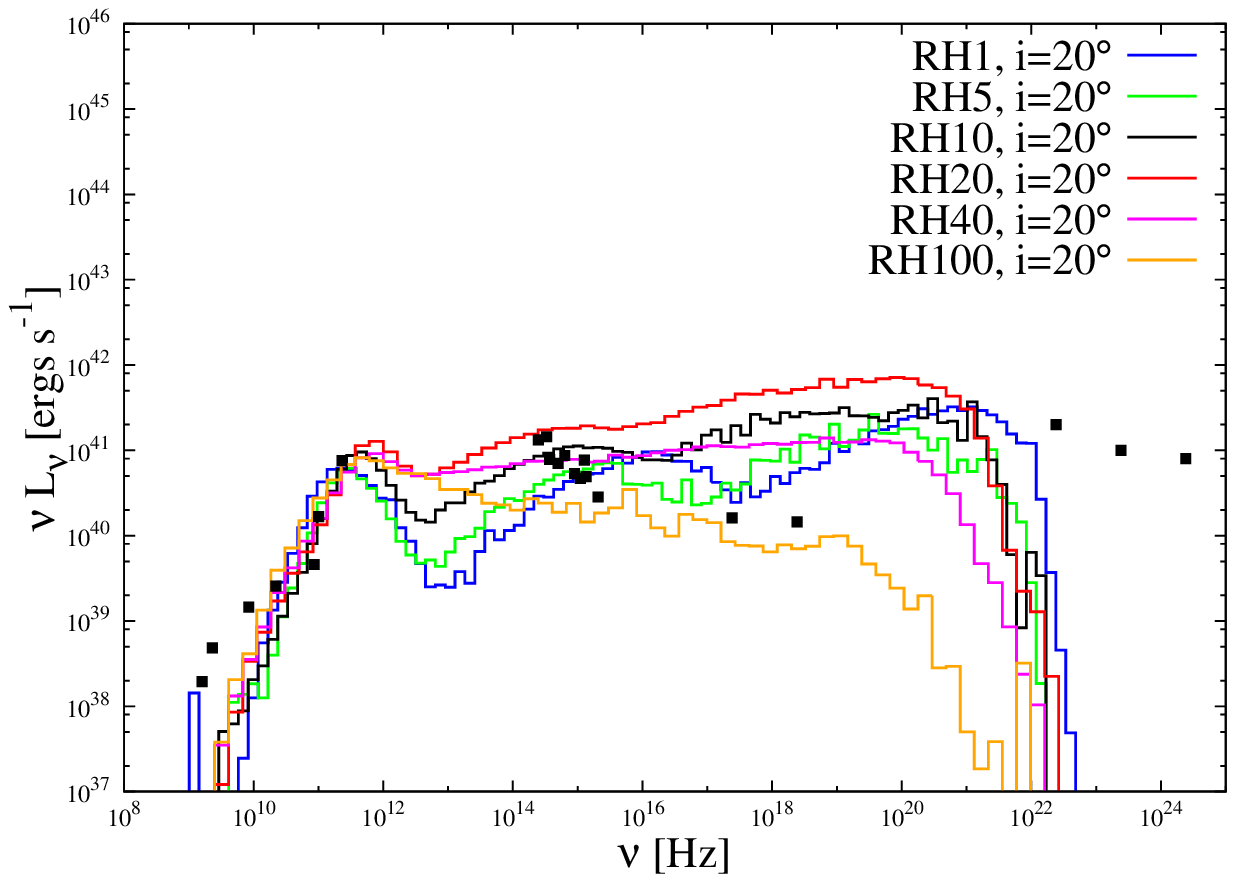}
\includegraphics[width=0.48\textwidth]{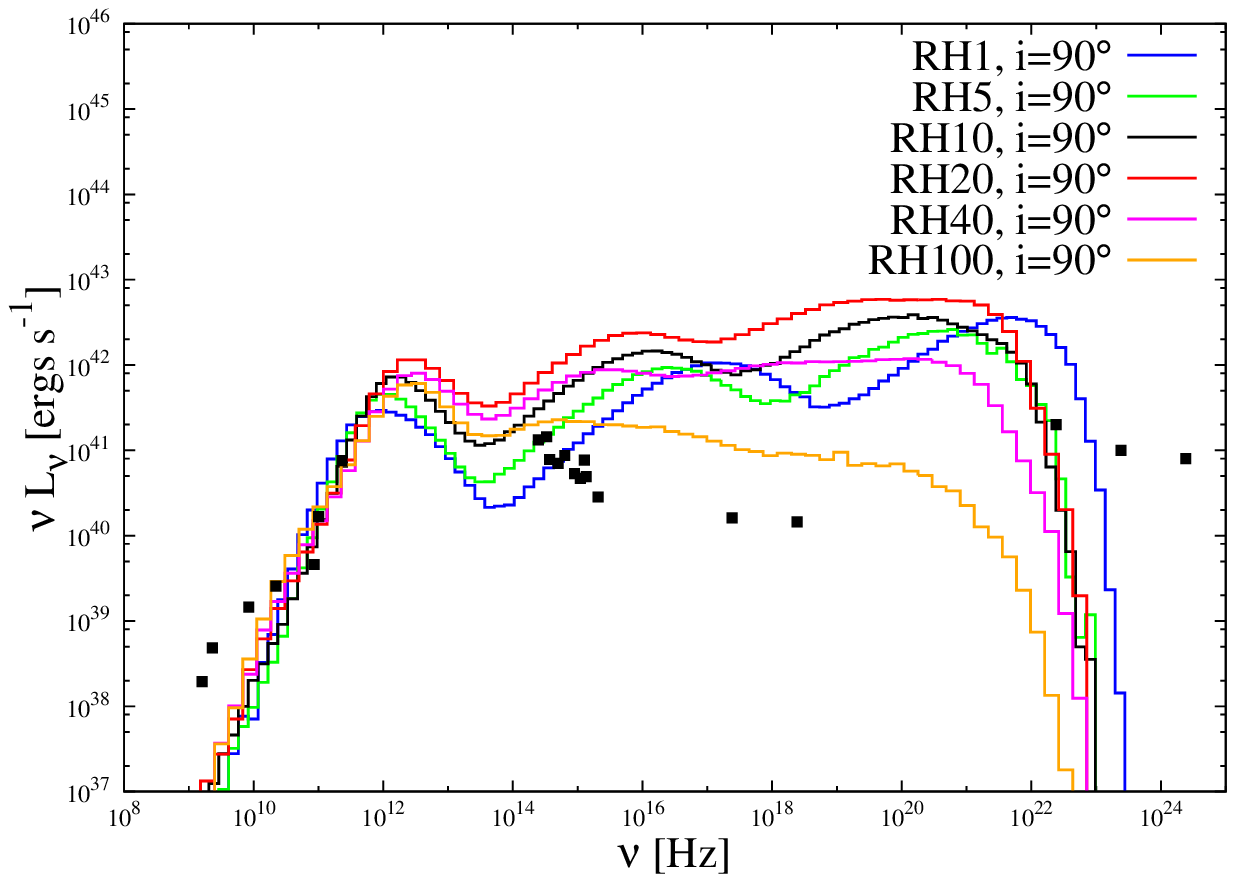}
\caption{Electromagnetic spectrum of GRMHD models computed at $i=20\degr$ and
  $i=90\degr$ based on models RH1-RH100 overplotted with the observations of
  M87 collected in \citet{abdo:2009}. 
Synchrotron emission appears in the SED as a hump around
230\,GHz, and the higher order humps (second and third ones) are  due to inverse-Compton emission.}\label{fig:seds}
\end{center}
\end{figure*}

Here, we present model spectra for six different combinations of $R_{\mathrm
  high}$ and $R_{\mathrm low}$ (in Eq.~\ref{eq:te}).
Table~\ref{tab:model_params} summarized the parameters used in these models.
Figure~\ref{fig:seds} shows SEDs of models RH1-RH100 for $i=20\degr$ and for
$i=90\degr$. The SEDs for $i=90\degr$ are given as reference to demonstrate
that most of the higher (than 230\,GHz) energy radiation is emitted from
the system in the direction (along the equatorial plane) away from our fixed
line of sight.  The numerical code includes radiative processes, such as the
synchrotron emission and self-absorption, which appears in the SED as a hump around
(230\,GHz) and the inverse-Compton scatterings (higher order humps). 

Another important  process is bremsstrahlung from electron-proton and relativistic electron-electron collisions.
We have implemented bremsstrahlung in our radiative transfer code (details
presented in Appendix D of
\citealt{moscibrodzka:2011}). Our calculations of bremsstrahlung emission from all
presented 3-D models indicate that the majority of the bremsstrahlung radiation
is produced in the accretion disk regions at radii $r > 24 \rg$ (beyond the
pressure maximum of the initial torus configuration). However, we note that the
accretion disk structure and dynamics at $r > 24 \rg$ are simply artifacts of the adopted initial plasma configuration (the torus). 
This makes any bremsstrahlung emission highly uncertain and
therefore omitted in the current work. Bremsstrahlung emitted from within
$24 \rg$ is negligible compared to the inverse-Compton emission.

In Fig.~\ref{fig:seds}, we also plot the observational data points from
\citet{abdo:2009} and \citet{doeleman:2012}. As already mentioned, our models
are normalized to reproduce the flux of 1 Jy at 1.3\,mm (230\,GHz). 
In Fig.~\ref{fig:seds}, the angular resolution of Fermi (data points at $E > 100 \mathrm{MeV}$ (or $10^{22}$ Hz) is large
($\theta \sim 0\degr.8\, E_{\mathrm{GeV}}^{-0.8}$), and data points include flux from
the entire jet including its radio lobes (e.g., include HST-1 and other knots
located further downstream of the jet, which are prime locations for the
particle acceleration, hence the high-energy emission). Therefore the
high-energy spectrum is used in this work as an upper limit.
The proton-to-electron temperature ratio in the accretion disk
has to be $R_{\mathrm high} \geq 100$ to produce a flux of 1 Jy at 1.3\,mm, but
not to overpredict the source luminosity measured at high energies.  

In the rest of the paper, we therefore focus on modeling images of the fiducial model
RH100 whose SED agrees with the observational data points. The radiative
efficiencies for all models and the importance of the radiative cooling is
discussed in \S~\ref{dis:cool}. The radio maps' appearance as a function of
$R_{high}$ is presented and briefly discussed in the Appendix.

\subsection{Emission at $\lambda=$7\,mm and 3.5\,mm}

\begin{figure*}
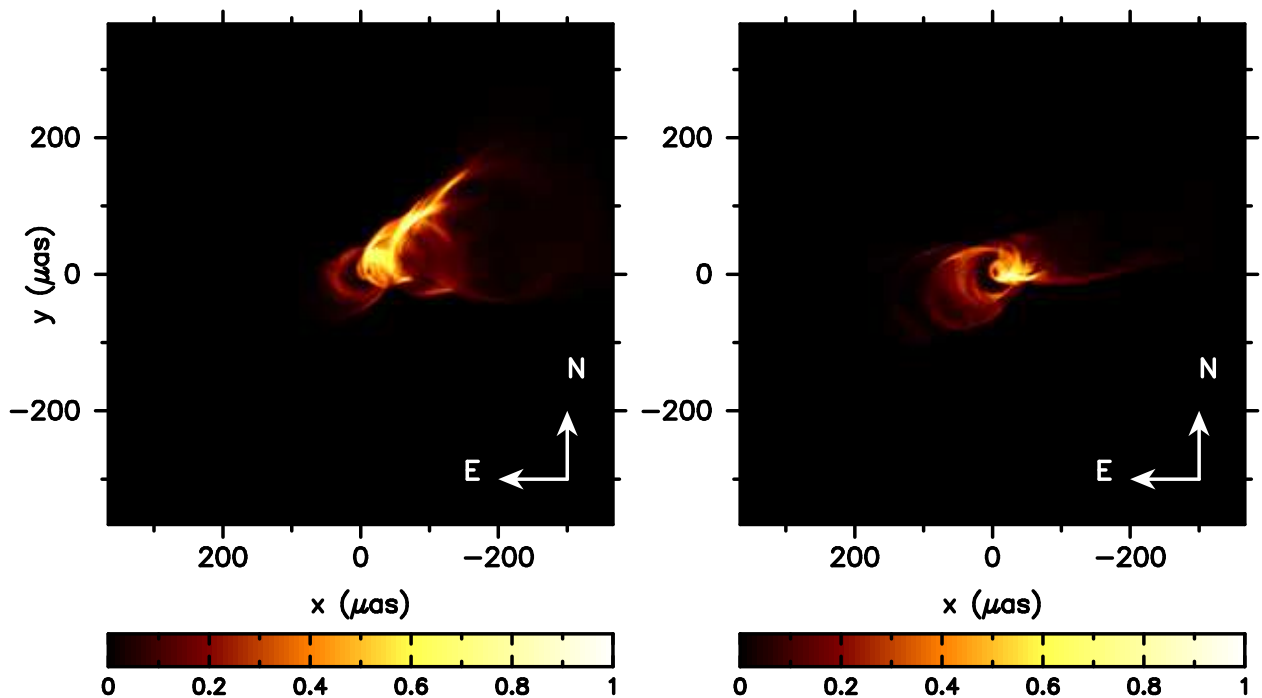

\begin{center}
\includegraphics[width=0.5\textwidth,angle=-90]{f4a.ps}
\includegraphics[width=0.5\textwidth,angle=-90]{f4b.ps}\\
\caption{Intensity map of our fiducial model RH100 at $\lambda$=7\,mm ($\nu$=43\,GHz) for
  a viewing angle of i=20\degr (left panel) and i=160\degr (right panel). 
  The color scale is normalized to unity. See 
  Table~\ref{tab:model_params} for the total fluxes in units of Jansky for
  each model (column 7). The position angle of the jet axis/black hole spin 
  is set to $PA=290\degr$ \citep{reid_1.6:1982} E of N for both models.
  The size of each panel is $200\times200\, \rg$ in the plane of
  the black hole. At a distance of D=16.7 Mpc, this corresponds to an angular size of
  about $0.8\times0.8$ mas.}\label{fig:img_43}
\end{center}
\end{figure*}

Figure~\ref{fig:img_43} shows the appearance of our fiducial model RH100 at $\lambda$=7\,mm
for $i=20\degr$ (left panel) and $i=160\degr$ (right panel).  The images show
the intensity distributions on the sky that are normalized to unity. They are
computed on the GRMHD model snapshot base and then time-averaged over a duration
of about 35\,d.  At 7\,mm, the plasma around the maximum of the intensity
distribution is optically thick (the synchrotron photosphere, $\tau_{\mathrm
  abs}=1$ surface, is located at a distance of about 10--25 $\rg$ from the
BH). The images display extended and complex structures that are evidently
edge-brightened. Moreover, there is the brightness asymmetry between the two
 rims on both sides of the jet.  The jet is corotating with the BH and
the disk. (The angular momentum vector of the BH, and the disk is pointing in
the N-W direction in the images.) The brightness assymetry is due to
Doppler boosting. In both cases shown in Fig.~\ref{fig:img_43}, the emission
from the counter-jet is strongly suppressed.

\begin{figure*}
\begin{center}
\includegraphics[width=0.45\textwidth,angle=0]{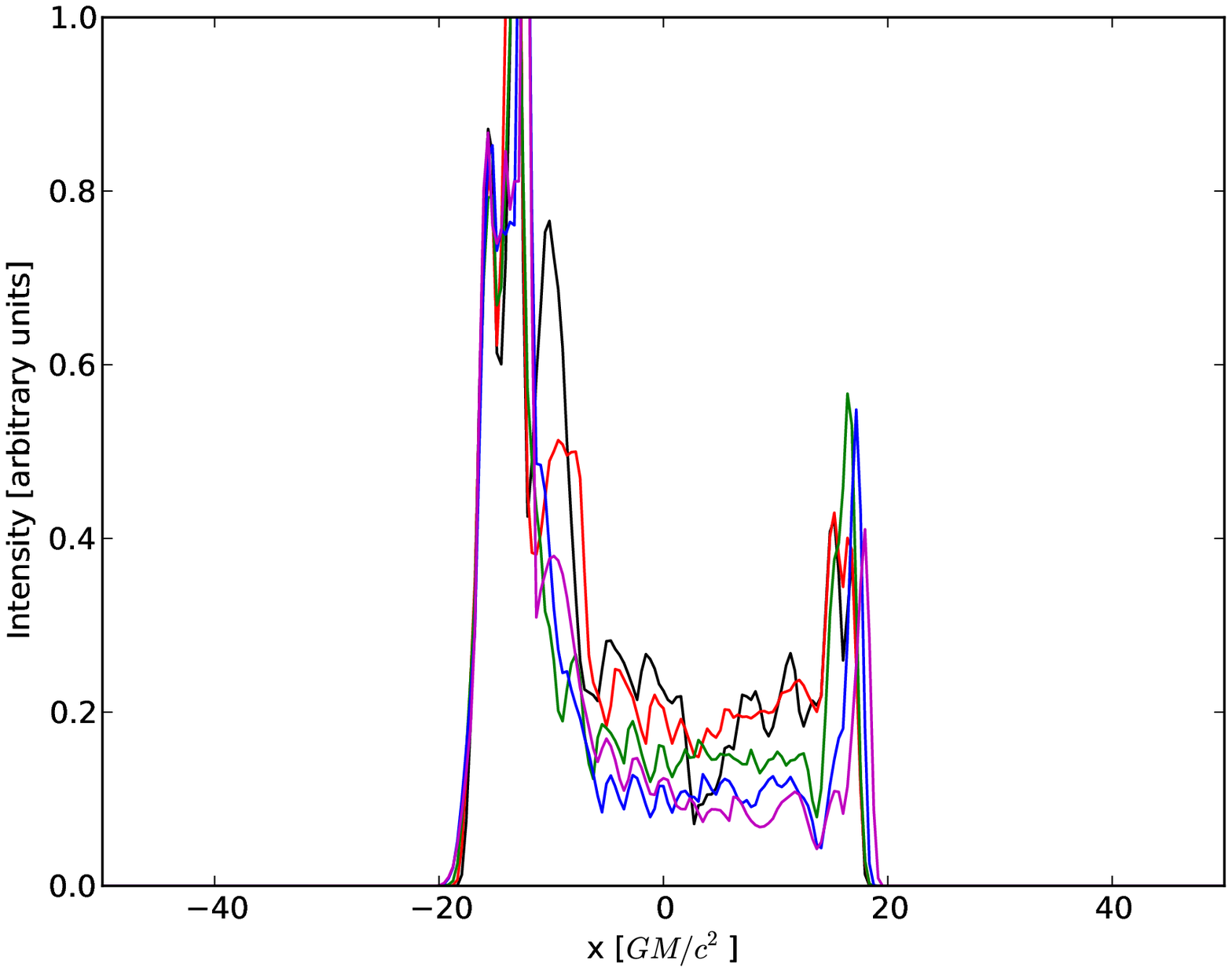}
\includegraphics[width=0.45\textwidth,angle=0]{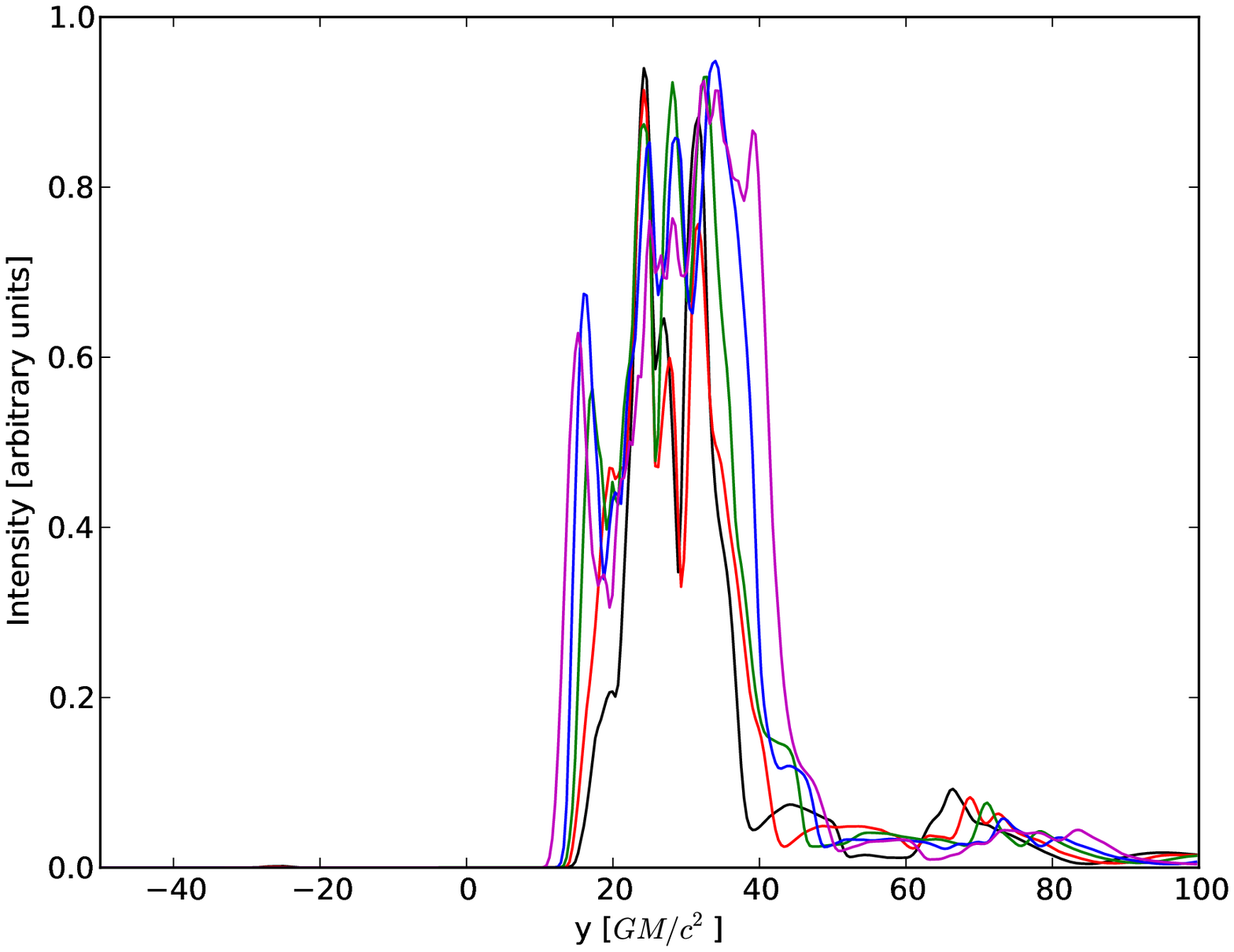}
\caption{Evolution of the intensity profile across (left panel) and along
  (right panel) the jet at $\lambda$=7\,mm ($\nu$=43\,GHz). Color lines represent intensity at various times
  spaced by $20M\equiv7$ days.}\label{fig:prof_43}
\end{center}
\end{figure*}

The edge-brightening of the jet images is illustrated in
Fig.~\ref{fig:prof_43} (left panel), which shows the radiation intensity
profile across the jet axis at a distance of $25\rg$ away from the SMBH.
Figure~\ref{fig:prof_43} (left panel) shows how the intensity profile evolves in
time. Lines with different colors indicate the intensity profile at various
times. The time span between the black and magenta lines corresponds to
about 28\,d. The ratio of the intensity of the two rims is about two and is
roughly constant in time.  Figure~\ref{fig:prof_43} (right panel) also shows the
evolution of intensity profile along the jet.  The profile along the jet shows
two intensity enhancements that apparently move upstream of the jet ("knots"
located at $x\sim45$ and $65 \rg$). We find that these two intensity "knots"
have subluminal apparent speeds of $v/c=0.13$ and 0.4, which indicate jet
acceleration.

\begin{figure*}
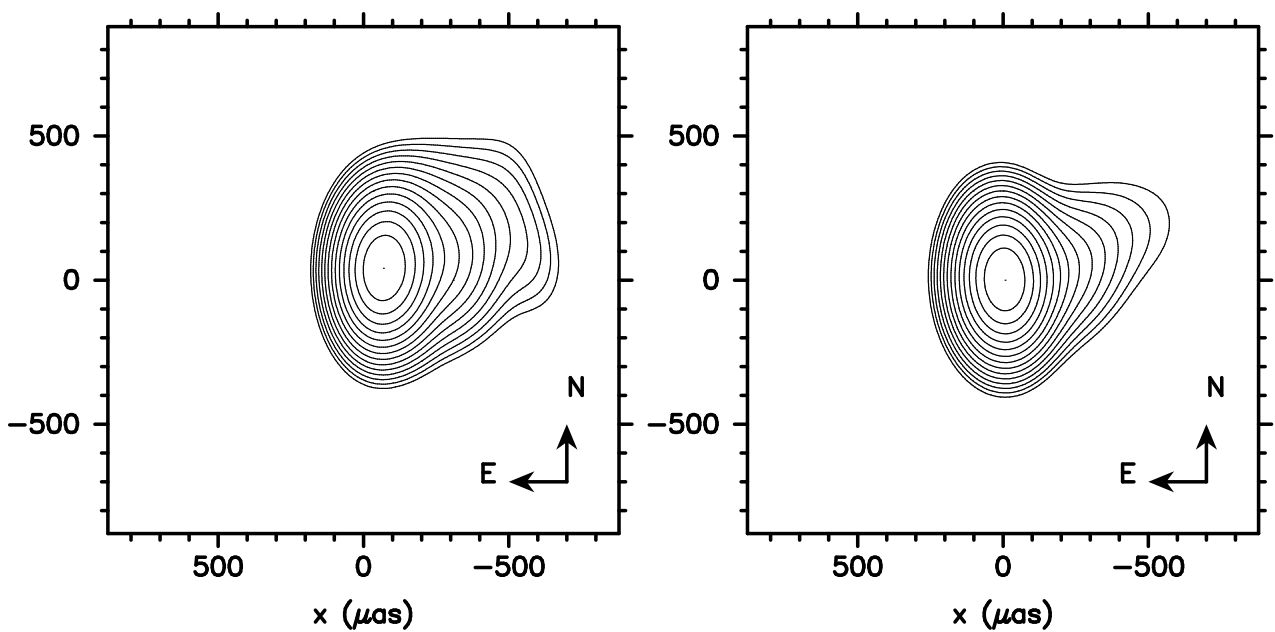

\begin{center}
\includegraphics[width=0.45\textwidth,angle=-90]{f6a.ps}
\includegraphics[width=0.45\textwidth,angle=-90]{f6b.ps}
\caption{Contour maps of the model images (RH100) at $\lambda$=7\,mm
  ($\nu$=43\,GHz) for viewing angles of i=20\degr (left panel) and i=160\degr
  (right panel) convolved with the telescope beam to simulate observations by
  \citet{hada:2011}.  The contour levels were chosen to match those from
  observations (contours decrease by a factor of $2^{1/2}$ from the maximum
  intensity).  The image size here is $480\times480 \rg \equiv
  1.8\times1.8$ mas, which is about twice the size used in
  Fig.~\ref{fig:img_43} (images at $\lambda=$7\,mm).}\label{fig:img_43_hada}
\end{center}
\end{figure*}

A robust comparison of Fig.~\ref{fig:img_43} to observations of the source at
7\,mm is presented in \citet{hada:2011}.  In Fig.~\ref{fig:img_43_hada}, we
convolve our theoretical intensity maps (Fig.~\ref{fig:img_43}) with the
telescope beam size ($FWHM_{\mathrm beam}$ =0.3 and 0.14 mas, see
\citealt{hada:2011}) and contour them in the same fashion as Fig.~3b in
\citet{hada:2011}. There is overall good qualitative agreement, but also some remaining differences.
Our jet model is somewhat more compact in the direction along
the jet axis to account for the extended low-surface brightness jet features observed at 7\,mm. Also
our jet model does not display the characteristic two rims when convolved with
the telescope beam, even though the underlying theoretical model is clearly edge-brightened.
An even better agreement between our model and observations
could probably be achieved (1) by using GRMHD models with a higher spatial
resolution that resolves the jet boundary better, (2) by increasing the size
of the computational domain since we are only simulating the innermost parts of
the jet at 43\,GHz, and (3) by including particle
acceleration mechanisms, i.e., power laws in electron DFs,
 which are omitted here for the sake of simplicity.
Adding a power law to the thermal energy distribution will amplify the
optical part of the emission, hence also enhance the extended jet
emission, which is optically thin after all.

\begin{figure*}
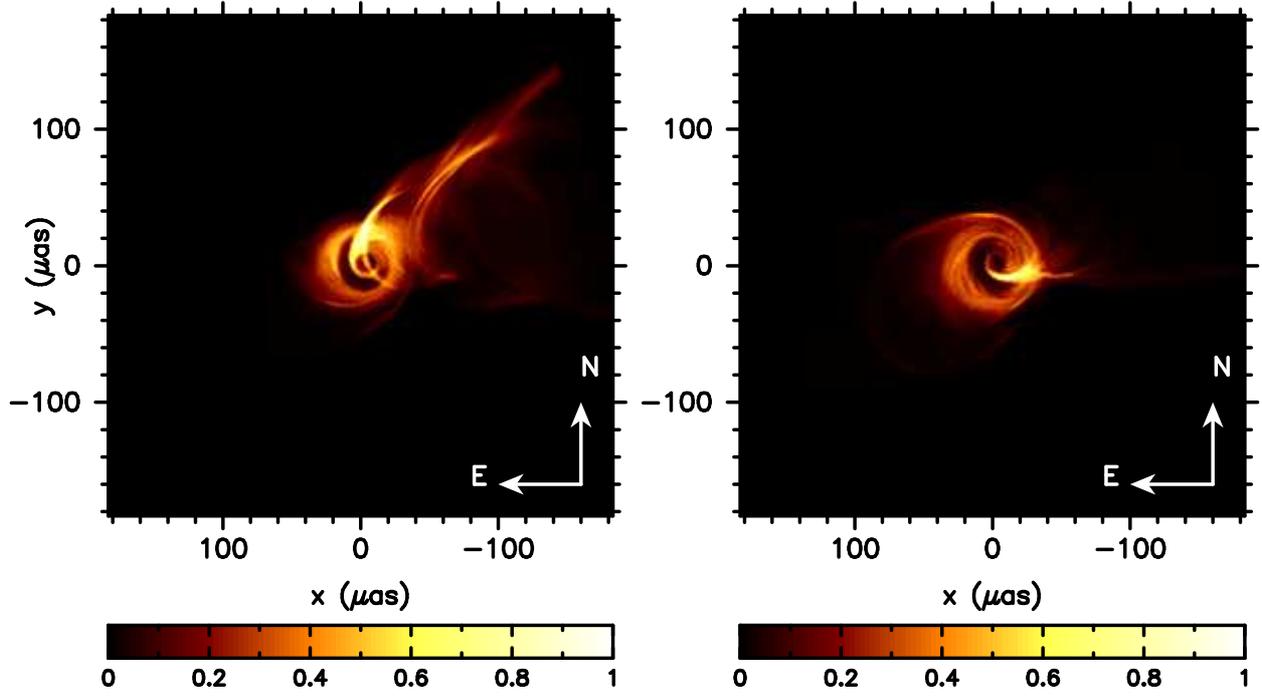

\begin{center}
\includegraphics[width=0.5\textwidth,angle=-90]{f7a.ps}
\includegraphics[width=0.5\textwidth,angle=-90]{f7b.ps}
\caption{Same as Fig.~\ref{fig:img_43} but for $\lambda$=3.5\,mm
  ($\nu$=86\,GHz). Here the panel size is $100\times100 \rg$ in the plane of the
  black hole, which corresponds to an angular size of about $480\times480
  {\mathrm \mu as}$.}\label{fig:img_3.5mm}
\end{center}
\end{figure*}

Figure~\ref{fig:img_3.5mm} shows two models' ($i=20\degr$ and $160\degr$) images
at 3.5\,mm. The emission from the jet has a parabolic shape (consistent
  with \citealt{hada:2011}), and it is
edge-brightened as in the 7\,mm images (Fig.~\ref{fig:img_43}).  The emission
from the counter-jet is more pronounced compared to the 7\,mm images.  However, owing
to lensing effects, the counter-jet images resemble a disk structure.  Overall, the emission becomes more compact
at $\lambda$=3.5\,mm compared to the one in
the 7\,mm image, indicating the well-known $\lambda$ dependency of the apparent sizes.

\subsection{Emission near the BH horizon at $\lambda=$1.3\,mm}

\begin{figure*}
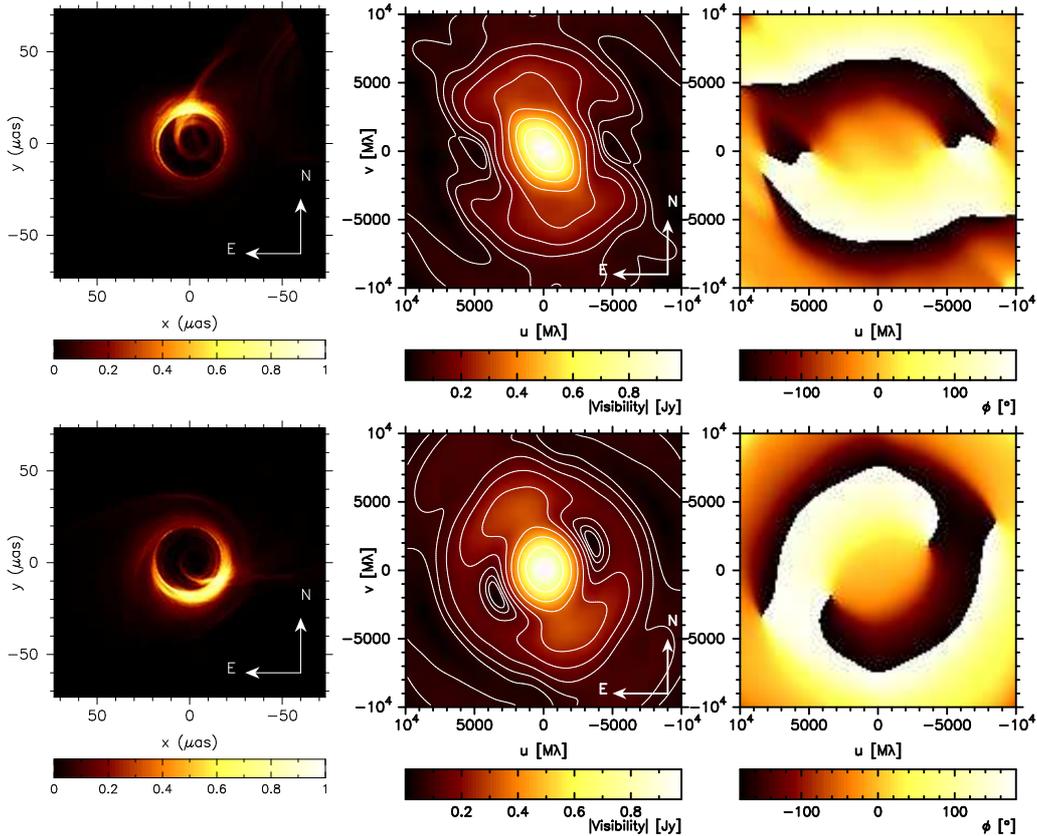

\begin{center}
\includegraphics[width=0.27\textwidth,angle=-90]{f8a.ps}
\includegraphics[width=0.3\textwidth,angle=-90]{f8b.ps}\\
\includegraphics[width=0.27\textwidth,angle=-90]{f8c.ps}
\includegraphics[width=0.3\textwidth,angle=-90]{f8d.ps}
\caption{Intensity maps of model RH100 for $i$=20\degr (upper left panel) and
  $i$=160\degr (lower left panel) at $\lambda$=1.3\,mm ($\nu$=230\,GHz).  The
  total fluxes (at 1.3\,mm) in these models are 1 Jansky.  The position angle
  of the black hole spin is set to $PA=290\degr$ E of N for all models.  The
  image size is $40\times40\,\rg$ in the plane of the black hole, which at a
  distance of D=16.7 Mpc, corresponds to an angular size of about $140\times140
  {\mathrm \mu as}$. Middle panels: the corresponding visibility amplitude on
  a u-v plane in units of Jy.  Right panels: the visibility phase map in
  degrees.  The arrows in the left and middle panels indicate the orientation
  of our coordinate system.}\label{fig:img_1.3mm}
\end{center}
\end{figure*}

In Fig.~\ref{fig:img_1.3mm} the lefthand panels display 1.3\,mm images of model RH100
for $i=20\degr$ and $i=160\degr$. At 1.3\,mm the images 
are dominated by the emission from the counter-jet, while
the emission from the jet on the near side (the one approaching the observer) is
notably weaker. The emission is produced near the jet
launching point and is compact. The emission around the BH has a shape of a crescent,
which is simply a lensed image of the counter-jet~\footnote{The emission from
an accretion disk will also produce crescent-like images when $R_{\mathrm high}$ is relatively low, see Appendix or, e.g., \citealt{moscibrodzka:2009},
\citealt{moscibrodzka:2014}, and \citealt{dexter:2012}.}. However, the approaching-jet appears as an
additional circle in front of the crescent. 
For our adopted position angle of the BH spin ($PA=290\degr$; \citealt{reid_1.6:1982}), the
emitting region is elongated in the E-W direction on the sky. In both images the BH
shadow size is about $40~\mathrm{\mu as}$, as expected. 

The core of M87 has been detected at 1.3\,mm with VLBI
\citep{doeleman:2012}. The reconstruction of a 1.3\,mm
radio map using these observations was not possible because of an insufficient number of 
mm-VLBI stations. In this section, the comparison of 1.3\,mm emission maps and the observations 
will be done in terms of interferometric observables.

The interferometric observables are the visibility amplitude 
and phase as a function of u-v coordinates (distances between pairs of millimeter telescopes or spacial frequencies).  
The complex visibility function that an interferometer "detects" 
(Fourier transformation of the intensity
distribution on the sky) is defined as follows
\begin{equation}
V(u,v)= \iint I(x,y) ~e^{-i(ux+vy)/\lambda} ~dx dy \equiv A ~e^{-i\phi}
\label{eq:vis}
\end{equation}
where $I(x,y)$ is the intensity distribution at a given set of lefthanded
coordinates on the sky (i.e., x and y are positive in E and N directions
on the sky, respectively), and $u,v$ are the projected lefthanded baseline
lengths (i.e., u and v are positive in eastern and northern directions on the Earth,
respectively). Our intensity
distributions based on GRMHD models are strongly non-Gaussian and so
$V(u,v)$ is a complex function that has non-trivial amplitude, $A$, and a non-zero phase,
$\phi$. In Fig.~\ref{fig:img_1.3mm} the middle and righthand panels show
the complex visibility amplitude and phase of the 1.3\,mm intensity maps (displayed
in the left panels), respectively.

\begin{figure*}
\begin{center}
\includegraphics[width=0.48\textwidth]{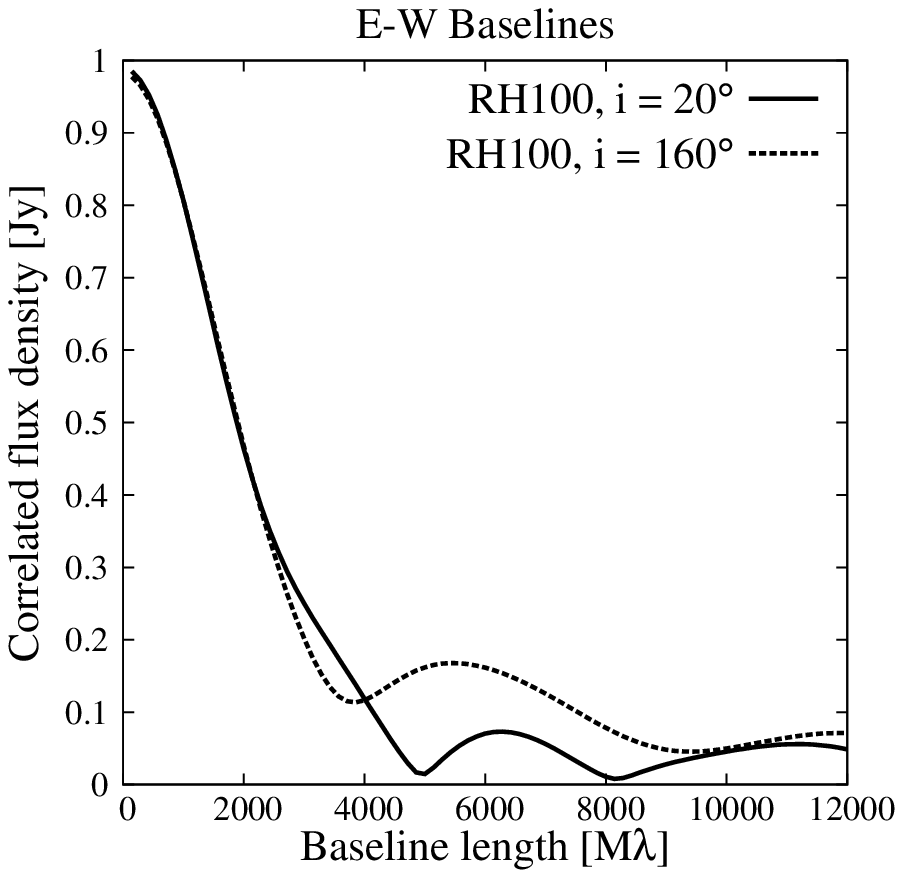}
\includegraphics[width=0.48\textwidth]{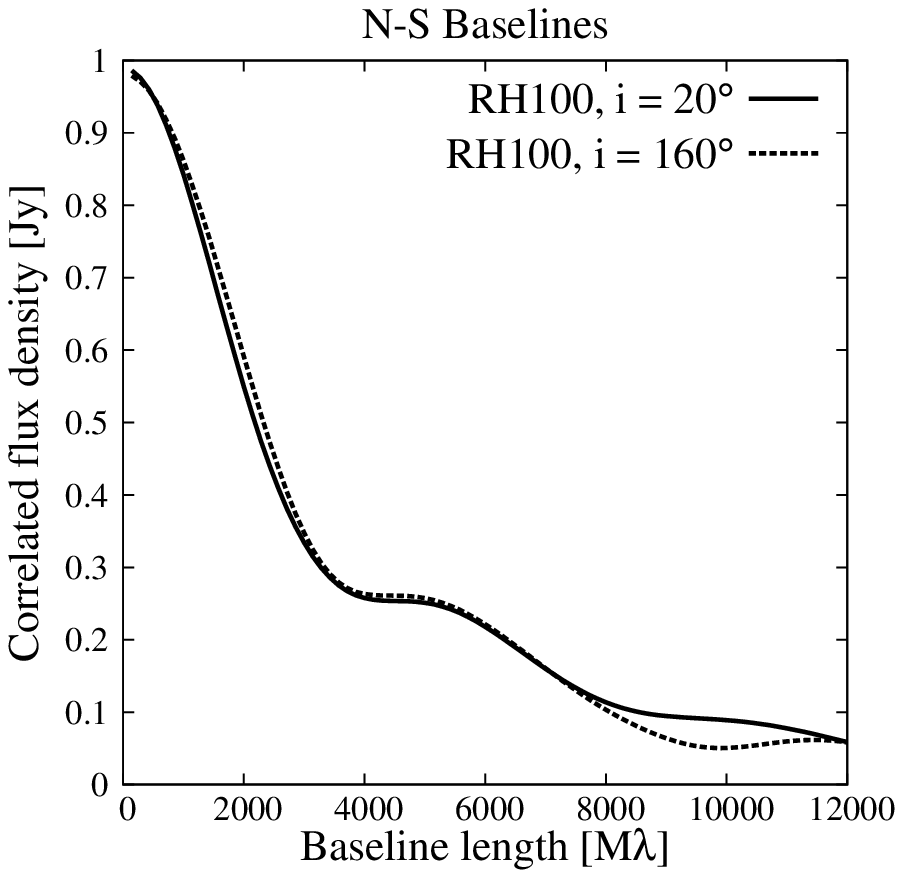}
\caption{Theoretical visibility amplitudes along E-W (left) and N-S (right)
  baselines, computed for model RH100 with i=20\degr (solid lines) and
  160\degr (dashed lines).}\label{fig:vis.baselines}
\end{center}
\end{figure*}

In the middle panel  of Fig.~\ref{fig:img_1.3mm}, the visibility
amplitude maps show two minima for baselines with sizes 5000--6500 ${\mathrm 
G}\lambda$, which correspond to the angular size of the BH shadow.
Figure~\ref{fig:vis.baselines} shows a cut of the visibility amplitude along E-W
and N-S baselines. Evidently, along E-W baselines, the model displays two
minima (at baselines of about about 4000--5000 and 10~000--8000
$\mathrm{M}\lambda$, where $M\lambda=$1.3 km) for both inclinations.  Along
N-S oriented interferometric baselines, there are no visible minima.

\begin{figure}
\begin{center}
\includegraphics[width=0.5\textwidth]{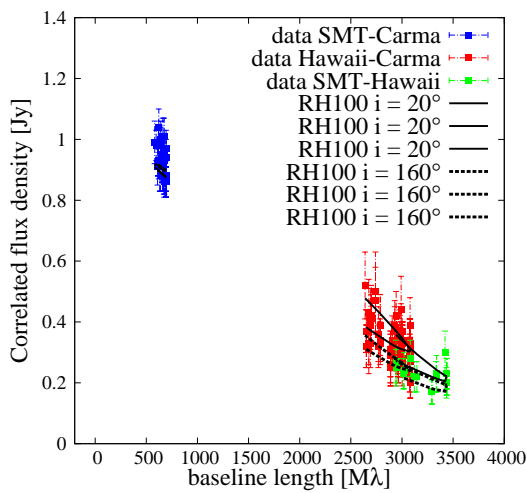}
\caption{Comparisons of the visibility amplitudes from our models and those
  from the EHT observations. The visibility amplitudes, computed at
  $\lambda$=1.3\,mm ($\nu$=230\,GHz), along three baselines (SMT-CARMA,
  SMT-HAWAII, and CARMA-HAWAII) are shown for model RH100 with i=20\degr
  (solid lines) and 160\degr (dashed lines).  The distance between the VLBI
  stations is expressed in units of ${\mathrm M}\lambda$ ($1
  \mathrm{M}\lambda\equiv$1.3 km). The observed visibility amplitudes (points)
  and the u-v tracks are taken from \citet{doeleman:2012}.}\label{fig:vis.data}
\end{center}
\end{figure}

The core emission of M87 has been detected at 1.3\,mm on VLBI baselines between the
Arizona Radio Observatory's Submillimeter Telescope (SMT), the Combined Array
for Research in Millimeter-wave Astronomy (CARMA), and the James Clerk Maxwell
Telescope in Hawaii (\citealt{doeleman:2012}). The baselines of the telescope
pairs are on the order of 800, 3500, and 4300 km (or equivalently 620, 2600,
and 3300 $\mathrm{M}\lambda$). In Fig.~\ref{fig:vis.data}, we make a crude
comparison of our model visibility amplitudes to those observed by the EHT to test whether
model RM100 is roughly consistent with the data for both inclination angles
($i=20\degr$ and $160\degr$).

The visibility phase and, in particular, the so-called {\it \emph{closure phase,}} which contains information on the
source structure, can also constrain the model. 
The closure phase is the sum of visibility phases for a triangle of interferometric baselines:
\begin{equation}
\phi_{closure}=\phi_{\mathrm SMT-CARMA}+\phi_{\mathrm CARMA-H}+\phi_{\mathrm H-SMT}.
\end{equation}
For a symmetrical Gaussian intensity distributions on the sky, the visibility phase 
is expected to be zero and so is the closure phase. Any deviation from a zero
closure phase will indicate the source deviation from a Gaussian or point-like
structure, and this observable can in principle be used to compare the model
and observed emission shape without reconstructing the radio maps.

We have calculated the theoretical visibility closure phases. 
For the model with $i=20\degr$, which shows the crescent on the N  side of the BH (see
Fig.~\ref{fig:img_1.3mm}), the closure phases are positive; $\phi_{\rm
  closure}$=11\degr,19\degr,11\degr,11\degr, where the four values correspond
to different time moments of the observation.  The $\phi_{\rm closure}$
evolution is caused by the rotation of the Earth, and it is probing slightly
different u-v values. A typical observation duration is ~two to three hours. This
is about three times shorter than the dynamical time scale of the source
(8.5\,h) with its BH mass, $6.2 \times 10^9 \msun$.  For $i=160\degr$, for
which the crescent is on the S side of the BH
(Fig.~\ref{fig:img_1.3mm}), the closure phases are negative: $\phi_{\rm
  closure}$=-21\degr,-21\degr,-12\degr,-9.5\degr.  In both cases, the values
are consistent with the observed value: $\phi_{\rm closure} \approx
\pm20\degr$ (Akiyama, priv. communications).

In summary, for the fiducial model RH100 (for both viewing angles of $i=20\degr$ and
$160\degr$), visibility amplitudes and closure phases are roughly consistent with the
preliminary observations of the M87 core obtained by the EHT (visibility
amplitudes and closure phases on a single VLBI triangle).

\section{Discussions}\label{sec:discussion}

Deriving the appearance of a jet in the direct vicinity of a SMBH is
not straightforward. The jet formation mechanism, as well as particle
acceleration in jets, is generally not understood well. Moreover, one
has to take spacetime curvature into account, which affects the plasma
dynamics and light propagation.  Using GRMHD simulations of a weakly
magnetized accretion flow, a jet appears naturally, and we calculate the
appearance of the M87 jet base at radio and millimeter
wavelengths. For the electron heating, we assume that the electrons
are weakly coupled to protons in the accretion disk and strongly
coupled in the jet -- a simple, but crucial concept that we have
already used successfully to explain the appearance of the SMBH in the
center of the Milky Way.  Below we discuss our results in a context of
observational constraints. We also discuss the model limitations.

\subsection{Mass-accretion rate}\label{dis:mdot}

The accretion rate onto M87 is estimated by fitting the GRMHD model SED to the
observed data points. The resulting best fit $\mdot$ will vary depending upon
the underlying electron distribution functions in the accretion disk and
jet. They typically vary between $\mdot=10^{-4}-10^{-2} \, \mdotu$
(\citealt{moscibrodzka:2011}, \citealt{hilburn:2012}, \citealt{dexter:2012}).
In our fiducial model RH100, we find $\mdot=9 \times 10^{-3} \, \mdotu$.

The mass-accretion rate can be also inferred from the observed Faraday
rotation (e.g., at 1.3\,mm). For example, \citet{kuo:2014} measured $RM = \pm
{\mathrm a few} \times 10^5 {\mathrm rad\,m^{-2}}$. They claim that this RM
puts an upper limit on the accretion rate of $\mdot_{\mathrm max} \approx 10^{-3}
\, \mdotu$.  However, the $\dot{M}$ derived from RM is strongly model
dependent. We therefore directly calculate the RM from our GRMHD models to test
whether our model agrees with the observed value by \citet{kuo:2014}.
The Faraday rotation measure is defined as
\begin{equation}\label{eq:far}
{\mathrm RM} = 10^{4} \frac{e^3}{2 \pi m_{\mathrm e}^2 c^4} \int f_{\mathrm ref} n_e B_{||} dl \, \, {\rm rad \, m^{-2}}
\end{equation}
where $\Theta_{\mathrm e}$, $n_{\mathrm e}$, and $B_{\mathrm ||}$ are the
dimensionless plasma electron temperature, density, and magnetic field
component projected along the null geodesics. For a relativistically hot
plasma, the correction term in Eq.~\ref{eq:far} becomes $f_{\mathrm rel} =
\log(\Theta_{\mathrm e})/2\Theta_{\mathrm e}^2$.  Density and field strength
in Eq.~\ref{eq:far} are given in c.g.s. units. 

The integration of RM is performed along the null geodesics from the observer
to the $\tau_{\lambda=1.3mm}=1$ surface.  The synchrotron photosphere at 1.3\,mm
is located near the event horizon, where the emission contribution is mainly
from the counter-jet.  For models RH1-RH100, the time-averaged and
intensity-weighted RM are $4 \times 10^{3}, 8 \times 10^{4}, -7\times10^{5},
-6\times10^{6}, -6\times10^{8}, and -1 \times 10^{10} \, {\mathrm rad \, m^{-2}}$,
respectively.  As a result, model RH10 matches the observational RM best.
Consequently, the mass-accretion rate of model RM10 is also
similar to the value estimate in \citet{kuo:2014}.  The RM in model RH100 is
too large. However, it is important to realize that the Faraday rotation
measured in our simulation is certainly overestimated and somewhat meaningless because of
the initial conditions of the torus (see, e.g., the density profile in
the accretion disk in Fig.~2 top left panel in \citealt{moscibrodzka:2014}).
This setup was chosen to provide a large mass reservoir from which the
SMBH is fed on small scales without the need for a continuous mass
flow from the outer boundary.

A reliable RM can therefore not be derived from our model, but
requires simulations that self-consistently feed the BH on large
scales. While this is computationally very expensive, this will be important to check in the future.

\subsection{Importance of radiative cooling}\label{dis:cool}

Our GRMHD simulation is decoupled from radiative transfer calculations.
To test that the radiation affects the dynamics of plasma in the simulation, we first examine
the radiative efficiency of our models. 
The radiative efficiency can be measured with a simplified formula: 
$\epsilon_{\mathrm r}=L_{\rm Bol}/\dot{M}c^2$, 
where $\dot{M}$ is the time-averaged mass accretion rate computed at the 
event horizon of the BH. 
The $\epsilon_{\mathrm r}$ are presented in Table~\ref{tab:model_params} (sixth column). 
One would hope that $\epsilon_{\mathrm r} < 1$, however, 
for most of the models (RH1-RH20) $\epsilon_{\mathrm r}>1$. These apparently too high radiative
efficiencies are unphysical and indicate either that models with RH1-RH20 are self-inconsistent 
or that the method of computing the radiative efficiency is oversimplified.
The high luminosities could be produced by a few localized places in the flow that contribute
to the total luminosity significantly, which might not be reflected in the
time-averaged mass accretion rate at the event horizon. We can either discard
the self-inconsistent models or propose more accurate ways of measuring radiative
efficiency of the accretion flow.

A more appropriate and accurate test for the importance of the radiative losses is to
compute the radiative cooling timescale $\tau_{\mathrm rad}=u/\Lambda$, where
$u$ is the specific energy density of gas and $\Lambda$ the cooling rate,
and compare it to the local dynamical timescale, $\tau_{\mathrm dyn} =
\sqrt{r^{3}/GM}$.  The basic radiative process is synchrotron cooling,
$\Lambda_{\mathrm syn}$.  Following \citet{moscibrodzka:2011}, we use
\begin{equation}
\Lambda_{\mathrm syn}= \frac{4 e^4 }{3 c^3
  m_e^2 }  n_e B^2 [\Theta_{\mathrm e}^{4/3} + (2\Theta_{\mathrm
    e})^{8/3}]^{3/4} \, \, erg\,cm^{-3} \, s^{-1}.
\end{equation}
We notice that $\Lambda_{\mathrm syn}$ should be reduced to account for the
synchrotron self-absorption and should be increased to account for the cooling
in the inverse-Compton process. The current $\Lambda_{\mathrm syn}$ is a crude
approximation of the real cooling function, which might be a few times larger.

Here we only use the simplified formula. Mimicking full RT (coupling
GRMHD with GRRT is beyond the scope of the present paper) could introduce
further uncertainties into the model. The uncertainties are due to the non-local,
multiwavelength, multidimensional, time-dependent nature of radiative transfer
equations. Using the simplified formula, in models RH1-RH10 we find the
synchrotron cooling rate to be shorter than the dynamical time scale only within
a few localized regions inside of $r=10\,\rg$. In other regions, it is ten times
and even longer than the dynamical timescale.  In these models, including radiation might indeed
cool the disk and, for example, increase the proton and
electron temperature contrast there.  
See \citet{moscibrodzka:2011} and \citet{dibi:2012} for examples. 
In models RH20-RH100, 
within 50 M, the cooling rate is a few to 1000 times longer than the dynamical time scale.
 Our fiducial model RH100 is therefore self-consistent.

In summary, we predict that including radiative cooling in the
GRMHD simulation will only cool electrons in the very inner disk. Owing
to uncertainty in the proton-electron coupling, the radiative cooling
might or might not influence the overall dynamics of the model. This
too needs to be investigated more thoroughly in the future, but it is also very compute intensive.
This does not, however, completely invalidate conclusions on the jet
  parameters.

\subsection{The jet power}

Another observational constraint on the M87 central engine is the
kinetic and magnetic power of the jet. Based on LOFAR observations, for example, the
large scale jet power is $P_{\rm j}=6-10 \times 10^{44} {\mathrm erg\,s^{-1}}$
\citep{gasperin:2012}. We notice that these jet power estimates are plasma model
dependent and that they are based on analysis of the emission from the halo surrounding 
M87. These observations can constrain our
model of the M87 core, but one has to note that the halo power is averaged 
over several million years, while the inner jet could change on much shorter
time scales. 
In the following we calculate the instantaneous jet power in our simulations.

The jet power is $P_{\rm j} =
\int_{\rm jet} \sqrt{-g} dx^2 dx^3 \left( F_{\rm E}^{(MA)}+F_{\rm E}^{(EM)}
\right)$, where $F_E^{(MA)}$ and $F_E^{(EM)}$ are dimensionless matter and
electromagnetic radial energy fluxes defined as $F_{\rm
  E}^{(MA)}=(T^r_t)^{(MA)} = (\rho_0 + u + p ) u^r u_t $ and $F_{\rm
  E}^{(EM)}=(T^r_t)^{(EM)} = b^2 u^r u_t - b^r b_t $, respectively. The
integration is done over the jet spine and sheath zones.  Since there is a
significant baryonic matter content in the jet sheath, the matter part of the
flux in the jet does not vanish. It does indeed dominate near the core. In
fiducial model RH100, the jet total power is $P_{\rm j} = 3 \times 10^{43} \,
{\mathrm erg\,s^{-1}}$, or $P_{\rm j} = 4 \times 10^{42} \, {\mathrm
  erg\,s^{-1}}$ if the rest mass flux is subtracted.

Our core jet power is therefore about 20--200 times lower than the
model-dependent average power needed to support the radio halo of M87.
On the other hand, our value is quite consistent with the jet power,
$P_{\rm j} \sim 10^{43} \, {\mathrm erg\,s^{-1}}$, derived by
\citet{reynolds:1996} from VLBI observations of the radio core itself.

As already mentioned, we studied models with a fixed BH
spin ($a_*\approx0.94$). Assuming that the jet power increases with
BH spin as $P_{\rm j} \sim a_*^2$ (\citealt{mckinney:2004},
\citealt{blandford:1977}), the discrepancy between halo and core jet
power cannot be removed even if an extremely fast-rotating BH ($a_*=0.998$) is assumed. The most likely explanation is
probably that the accretion rate is much lower now than it was a few
million years ago. Alternatively, the jet could be way out of
equipartition by transporting much more energy in either protons or
magnetic fields and having a much lower radiative efficiency, but this
would also require a much higher accretion rate and produce more
Faraday rotation.

\subsection{Equation of state}

In the GRMHD simulation, we have used the gamma-law equation of state (EOS)
with the adiabatic index $\gamma_{\rm ad}=13/9$. This value is appropriate for
a gas with relativistic electrons with $T_{\mathrm e}~>6 \times 10^9$ and
non-relativistic protons with $T_{\mathrm p} < 10^{13} \,\mathrm{K}$, i.e.,
assuming that their temperature ratio is fixed to 1 (Shapiro 1973).  For model RH100, this is only valid in the jet zones. In the disk, an EOS with
$\gamma_{\rm ad}=$5/3 should be used instead. However, simulations analogous to
our model but with different adiabatic indices do not show any noticeable
difference in their dynamics \citep{shiokawa:2013}.

\subsection{Predictions for 1.3\,mm VLBI observations and detecting the black hole shadow}

Our 1.3\,mm images are dominated by emission from the counter-jet (see
Fig.\ref{fig:img_1.3mm}).  If model RH100 is reasonable, we expect the BH shadow to be detected by the EHT with a baseline in the E-W direction and
a length of about 6500 km. There should be a second minimum in the visibility
around the 10400 km baseline, also along the E-W direction. The possible mm-VLBI
stations involved in the detection of the BH shadow would then be Hawaii-LMT
(first minimum) and PV-CARMA or PV-Arizona (second minimum). Moreover, here we
find that the exact position of the visibility minimum could be sensitive to
the direction of rotation of the jet (and the BH spin) on the sky.

Our conclusions about the detectability of the BH shadow are consistent with those
of \citet{dexter:2012},  who focus on modeling the emission from the core of
M87 emission using their 3-D GRMHD model. Their electron DF is a power-law
function, and the electron acceleration efficiency is a function of plasma
magnetization; i.e., the efficiency is stronger in highly magnetized jet
zones. Such assumptions also make the jet brighter.

\citet{broderick:2009} have presented semi-analytical GRMHD models to fit SED,
radio, and mm images of the M87 radio core. Their study is more comprehensive because
they include modeling of polarized synchrotron emission.  Most of their models
have an extremely fast-rotating BH ($a_*=0.998$), but in this case the BH
shadow at millimeter wavelengths is also illuminated by the counter-jet, which is
geometrically more extended compared to our GRMHD numerical
models. \citet{broderick:2009} did not restrict the orientation of appropriate
baselines to E-W direction in their analysis because their counter-jet emission
is more uniform and extended compared to ours. At 7\,mm, the size of the jet
emission in their model is sensitive to the jet collimation parameter, $\xi$.
It is quite possible that our model will also be sensitive to the exact value of the BH spin or the disk size that may control the jet collimation. We leave this issue
for a future study.

\subsection{Matter content of the jet funnel}

Unlike those by \citet{broderick:2009} and
\citet{dexter:2012}, our model does not make any assumption about the jet matter
content.  At the jet wall (sheath), the plasma is baryonic, and it is
constantly supplied from the accretion disk. Inside of the jet funnel,
electron-positron pairs are produced in $\gamma\gamma$ collisions.  We
calculate the pair production rate directly from the RT model following
\citet{moscibrodzka:2011}. In model RH100, the pair production rate near the
SMBH horizon is $\dot{n}_\pm=7 \times 10^{-4}$ (based on Eq.~26 in
\citealt{moscibrodzka:2011}, assuming that the luminosity of the source around the
electron rest-mass energies is $L_{512 keV,M87}\approx 10^{41} {\mathrm
  erg\,s^{-1}}$). This gives an estimate for the pair plasma density near the
event horizon $n_{\pm}\approx \dot{n}_\pm \tu_{M87} \approx 21
\,\mathrm{cm^{-3}}$.  Using Eq.~45 in \citet{moscibrodzka:2011}, we can
calculate the Goldreich–Julian density $n_{GJ}$, i.e.~the pair density
required to enforce the ideal MHD condition (E = 0) in the rest frame of the
plasma \citep{gj:1969}. We find that $n_{\mathrm{GJ}}\approx 5 \times 10^{-6}
{\mathrm{cm^{-3}}}$ in model RH100; i.e., the pairs cannot be produced in
cascades and multiply themselves. In summary, the pair density in the jet
funnel (spine) remains low.

\subsection{Emission from the jet sheath in radio}

\begin{figure*}
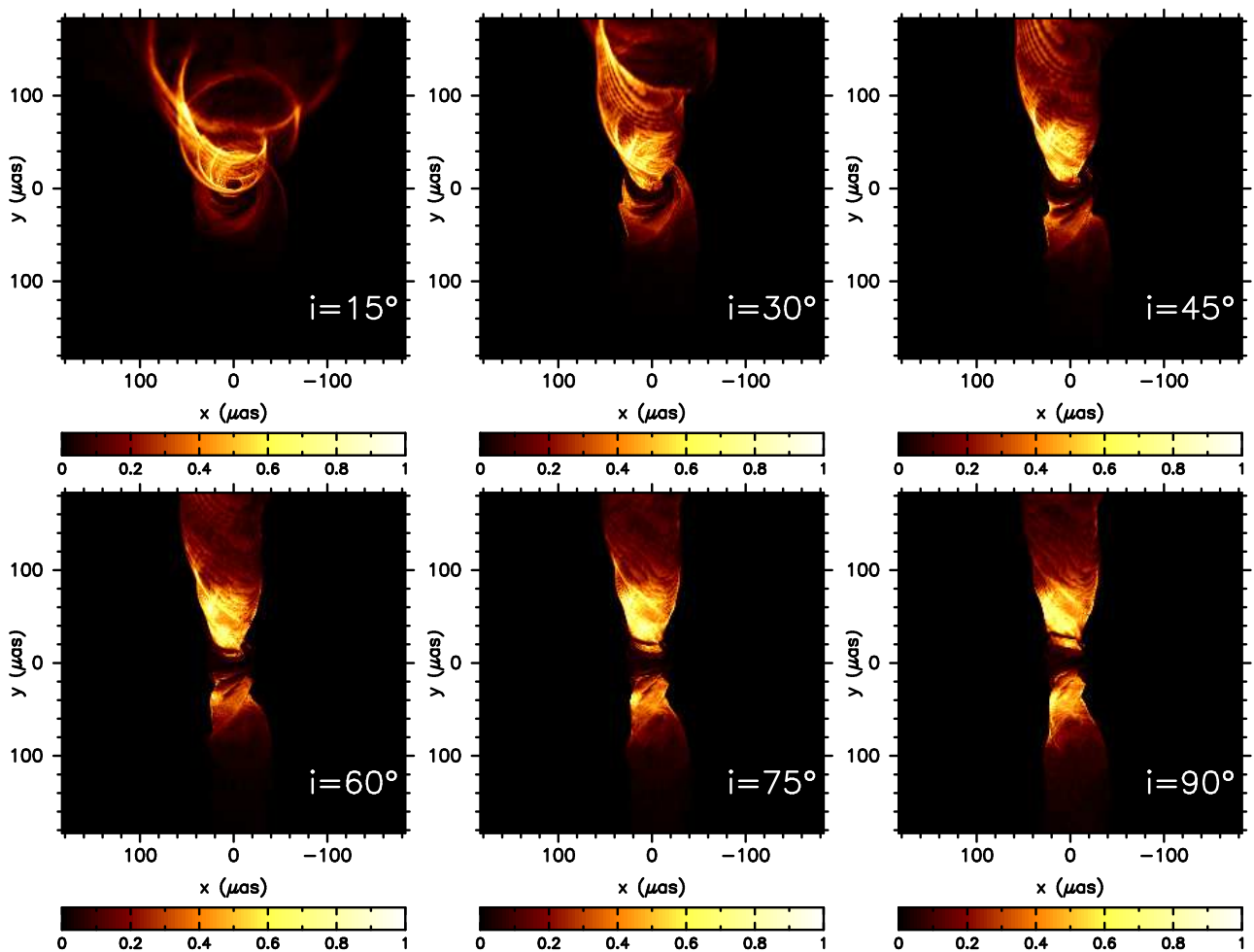

\begin{center}
\includegraphics[width=0.35\textwidth,angle=-90]{f11a.ps}
\includegraphics[width=0.35\textwidth,angle=-90]{f11b.ps}
\includegraphics[width=0.35\textwidth,angle=-90]{f11c.ps}\\
\includegraphics[width=0.35\textwidth,angle=-90]{f11d.ps}
\includegraphics[width=0.35\textwidth,angle=-90]{f11e.ps}
\includegraphics[width=0.35\textwidth,angle=-90]{f11f.ps}
\caption{43\,GHz images of model RH100 observed at various inclination angles.}\label{fig:inclination}
\end{center}
\end{figure*}

Finally, we briefly discuss the effect of the jet vertical stratification (jet
spine and sheath) on the observational properties of jets in general. A recent
work by \citet{boccardi:2015} reports the detection of the edge-brightened
jet in Cyg~A.  The jet in Cyg~A jet is not resolved by radio observations as
well as is the M87 jet. Therefore we should keep in mind that their scales are
different and their inclination angles are also probably different.  The authors
report that the counter-jet is narrower than the approaching jet. To
clarify what our models predict we plot our 43\,GHz images of the M87 jet at various
inclination angles in Fig.~\ref{fig:inclination}.  Here we do not notice any
evident difference in the opening angle of the jet and counter-jet owing to the
geometrical effects of light propagation.  However, we find that the jet
edge-brightening depends on inclination, and it becomes more notable for
$i\lesssim30\degr$.

\begin{figure*}
\begin{center}
\includegraphics[width=0.45\textwidth]{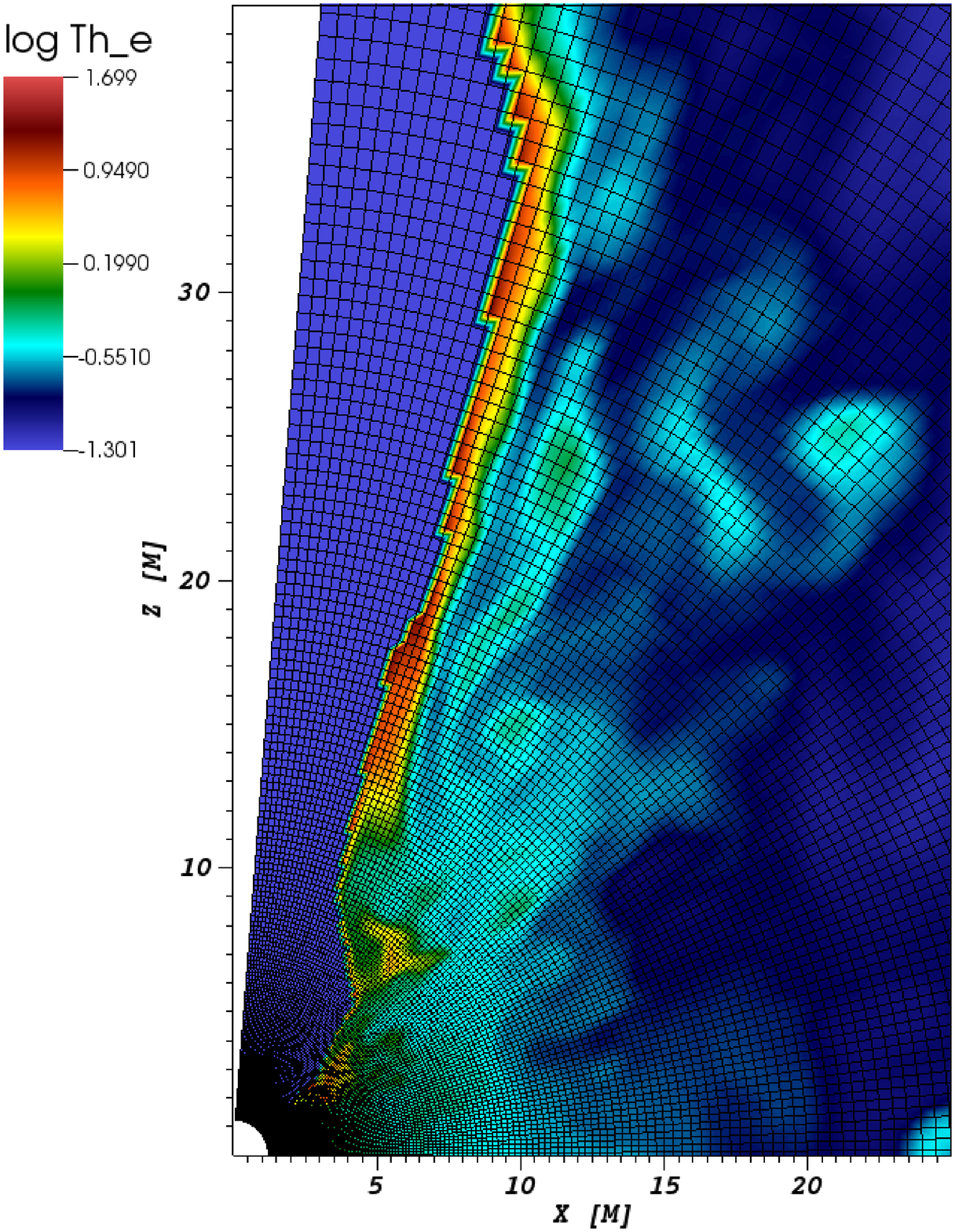}
\caption{Dimensionless electron temperature, $\Theta_{\rm e}=kT_e/m_ec^2$, in
  model RH100 overplotted with numerical grid. Maps show a cut through the 3-D model along the
  BH spin axis.}\label{fig:thetae_grid}
\end{center}
\end{figure*}

In the images presented in Sect.~\ref{sec:results},
the signal-to-noise ratio is greatly improved thanks to the time-averaging of 
many frames. However, we notice that single images of the jet at longer wavelengths
exhibit some numerical artifacts. These are most visible 
in the upper panels in Fig.~\ref{fig:inclination}.
The influence of the jet-wall resolution on the emitted radiation should be
addressed in the future. Figure~\ref{fig:thetae_grid} shows that in our
simulation, the jet wall is indeed poorly resolved numerically.

\section{Conclusions}\label{sec:conclusions}

We find that our radiative GRMHD simulations can account for many of
the observational characteristics of the M87 radio core and jet, such
as the edge-brightening, the size-wavelength relation, and the subluminal
apparent motion of the jet near the BH. The jet in the GRMHD
simulations is naturally produced if a poloidal magnetic field is
accreted. The sheath surrounding this jet carries most of the energy
and is best identified with the jets observed in radio observations.

In fact, the jet sheath (or ``funnel wall'') we find in the GRMHD
simulations recovers the basic properties of flat-spectrum radio cores
as described by the simple analytic solutions for the jets of \citet{blandford:1979}
and \citet{falcke:1995} and hence may be of
interest for flat spectrum radio cores in general. A noticeable
requirement to explain the jet dominance of the accreting system is --
apart from relativistic beaming -- the need to have cooler
two-temperature plasma in the accretion flow and hot proton-electron
plasma in the jet. Obviously, had jet and disk the same electron
temperature, the latter would always dominate in a coupled jet-disk
system since it simply contains more particles. Electron-positron
pairs in the jet are neither needed nor naturally produced in this
picture.

The current model is consistent with the size of the M87 core measured
at 1.3\,mm with VLBI on three baselines \citep{doeleman:2012}.  Our
best-fit model reveals that the emission at 1.3\,mm is also likely
to be produced by the plasma in the jet and not by the accretion disk.  Our
model suggests that the BH is in fact illuminated by the counter-jet
at 1.3\,mm. Owing to strong gravitational lensing effects, the image of the
counter-jet has a shape of a crescent, therefore the BH shadow could
be only detectable on interferometric baselines oriented in certain
directions. For position angles of the jet axis (which coincide with
the BH spin axis) at PA=290\degr, the source at 1.3\,mm is
expected to be elongated in the E-W direction.  Based on our current
models, we conjecture that the minimum in the visibility amplitude,
which corresponds to the shadow of the BH, could be observed
along the E-W interferometric baselines of the EHT.

Our jet models are still a bit too compact to accurately match the extended
and large-scale radio images of M87 at 3.5 and 7\,mm.  We speculate that
one reason could be the neglect of non-thermal power-law electrons in
our calculations. Better agreements between our models and observed
jet image sizes and power could probably be achieved by (1) adopting different
values of the BH spin, (2) including additional particle
acceleration mechanisms, and (3) using a higher grid resolution that can
resolve the jet wall better.

It is interesting to note that our GRMHD model was initially developed
for the Galactic center SMBH Sgr A* and scaled to M87. This seems to
work in a relatively straightforward manner despite many orders of
magnitude difference in accretion rate and mass. M87 may need a
somewhat hotter jet than Sgr A*. The main properties of the M87 jet
can be reproduced when we assume that the proton-to-electron temperature ratio
in the relatively weakly magnetized (advection-dominated) accretion disk is rather large ($\trat=100$), while 
in the jet, one has a single temperature ($\trat=1$). The high temperature
ratio in the disk was postulated in the early advection-dominated
accretion flow models by \citet{narayan:1995}, and the potential
difference in $\trat$ between jet and disk was later suggested by
\citet{yuan:2002}. As a result, we are not presenting radically new
  ideas. However, combining both ideas and integrating them into large
  numerical simulations seems to represent an important  step forward toward
  understanding the appearance of jets.

In the future, the following issues should be addressed: the grid resolution
in the jet wall, the electron acceleration along the jet wall, dependence of
the jet radiative properties on a BH spin, and feeding BH from
self-consistent boundary conditions instead of a torus.

\begin{acknowledgements}
This work is supported by ERC Synergy Grant ``BlackHoleCam: Imaging the Event
Horizon of Black Holes'' awarded to Heino Falcke, Michael Kramer, and Luciano
Rezzolla.  We would like to acknowledge Scott Noble for providing HARM-3D, which
was used to run 3D-GRMHD model. We thank Charles F. Gammie for providing
computational resources to run the GRMHD model on XSEDE (supported by NSF
grant ACI-1053575).  We thank Charles F. Gammie, Jonathan McKinney and 
Ryuichi Kurosawa for their comments.

\end{acknowledgements}

%\bibliographystyle{aa}
%\bibliography{local}

\appendix

\section{Radio emission at 3.5 and 1.3\,mm as a function of model parameters}

\begin{figure*}
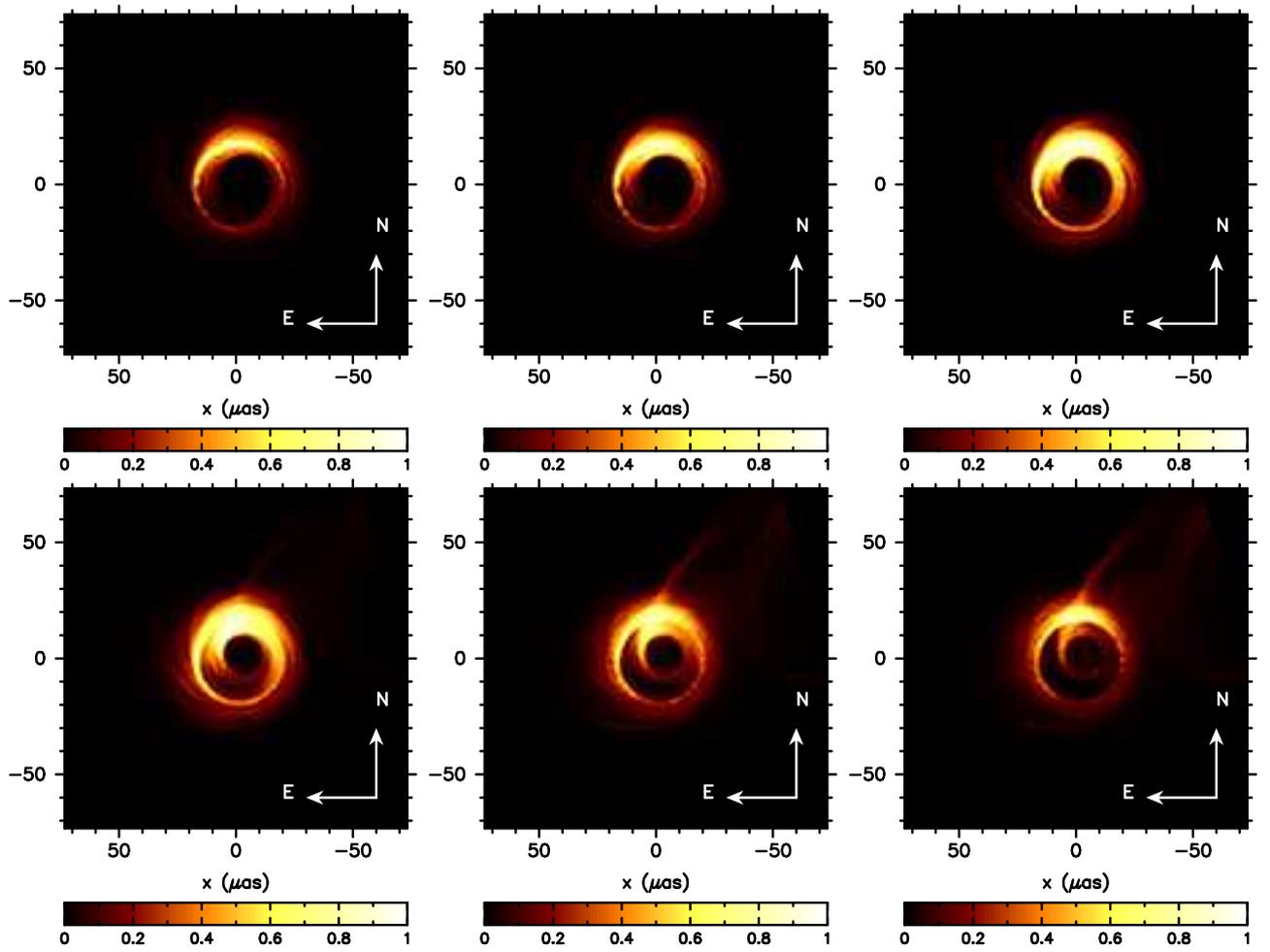

\begin{center}
\includegraphics[width=0.35\textwidth,angle=-90]{Af2.ps}
\includegraphics[width=0.35\textwidth,angle=-90]{Bf2.ps}
\includegraphics[width=0.35\textwidth,angle=-90]{Cf2.ps}
\includegraphics[width=0.35\textwidth,angle=-90]{Df2.ps}
\includegraphics[width=0.35\textwidth,angle=-90]{Ef2.ps}
\includegraphics[width=0.35\textwidth,angle=-90]{Ff2.ps}
\caption{Radio maps of models RH1-100 observed at $\lambda$=1.3\,mm ($\nu$=230\,GHz).
Upper panels from left to right are models RH1, RH5, RH10 and 
lower panels from left to right are models RH20, RH40, RH100.}\label{fig:all230}
\end{center}
\end{figure*}

\begin{figure*}
\begin{center}
\includegraphics[width=0.35\textwidth,angle=-90]{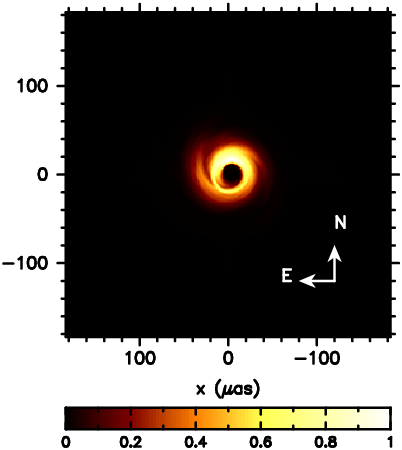}
\includegraphics[width=0.35\textwidth,angle=-90]{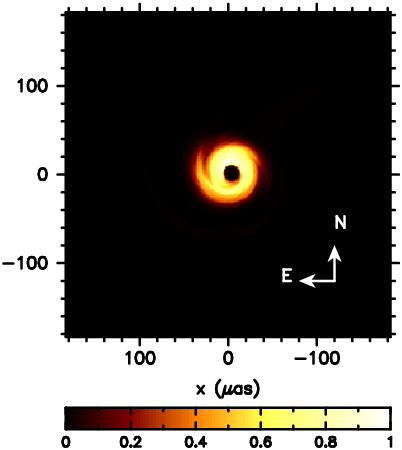}
\includegraphics[width=0.35\textwidth,angle=-90]{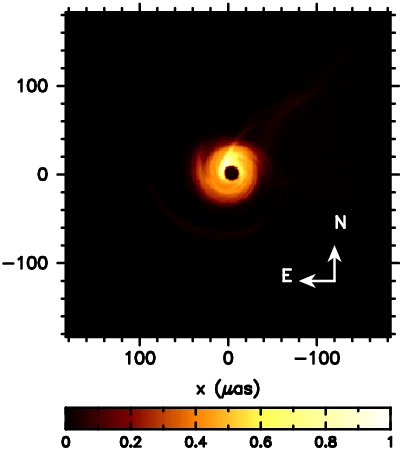}
\includegraphics[width=0.35\textwidth,angle=-90]{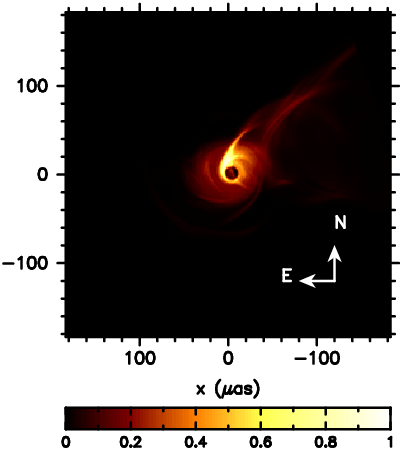}
\includegraphics[width=0.35\textwidth,angle=-90]{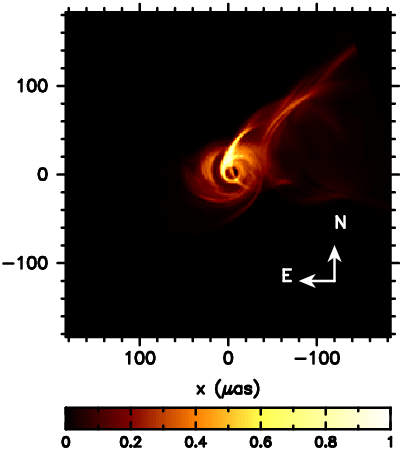}
\includegraphics[width=0.35\textwidth,angle=-90]{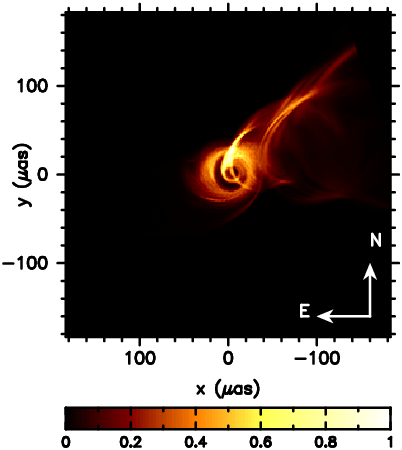}
\caption{Radio maps of models RH1-100 observed at $\lambda=3.5$\,mm ($\nu=86$\,GHz).
Upper panels from left to right are models RH1, RH5, RH10 and 
lower panels from left to right are models RH20, RH40, RH100. 
}\label{fig:all86}
\end{center}
\end{figure*}

Figures~\ref{fig:all230} and \ref{fig:all86} show model RH1-RH100 intensity
maps at 1.3 and 3.5\,mm, respectively. The color scales on each
panel are scaled linearly normalized to unity to clearly show the dynamical range of the
radio maps. The total flux at 1.3\,mm images is always one Jansky (since
all models are normalized to reproduce it), and total fluxes for 86\,GHz emission are provided in
Table~\ref{tab:model_params}. All images have been rotated so that the BH
spin has a positional angle on the sky oriented at PA=290\degr. 

The strongly Doppler-beamed and gravitationally lensed 1.3\,mm (230\,GHz)
images show crescents and rings in all cases. Models with low
$R_{\mathrm{high}}$ are dominated at this frequency by emission from the very
inner parts of the accretion disk (within the ISCO). Models with higher
$R_{\mathrm{high}}$ are dominated by the counter-jet emission. Models with higher 
values of $R_{\mathrm{high}}$ (e.g., RH20, RH40, and RH100) 
also display additional features that are produced by the jet wall on the near side, and
the feature shape resembles the shape of the number `6'. The shape of the
near-side jet emission is due to the combined effects of the parabolic jet-wall
shape and Doppler boosting. 

At a longer wavelengths 3.5\,mm ($\nu=86$\,GHz), the difference between models with 
low (RH1-10) and high $R_{\mathrm{high}}$ (RH20-100) becomes significant.
For low $R_{\mathrm{high}}$ (upper panels in Fig.~\ref{fig:all86}), the radio
emission is primarily produced by the accretion disk and so the Doppler
boosting and gravitational lensing are weaker.
For high values of $R_{\mathrm{high}}$
(lower panels), the radio maps are dominated by the emission from jets.
The BH shadow (black circle in the center of all maps) 
is visible in all 86\,GHz radio maps, and its size is apparently changing
because of obscuration effects. The plasma in front of the BH is optically thick
and is partially blocking our view. In models with higher values of
$R_{\mathrm{high}}$, the emission is more optically thin, and the BH shadow 
has its expected size of about 40 $\mu as$.

\end{document}